\documentclass[twocolumn,9pt]{article} 

\usepackage[square,numbers,sort&compress,comma]{natbib}

\usepackage{amsmath}
\usepackage{amssymb}
\usepackage{caption}
\usepackage{booktabs}
\usepackage{graphicx}
\usepackage{latexsym}
\usepackage{times}
\usepackage{siunitx}
\usepackage[pagewise]{lineno}
\usepackage{hyperref}

\topmargin - 12pt 
\renewenvironment{abstract}%
              {
               \small
               {\bfseries \abstractname}
               \par
               \vspace{10pt}
              }

\renewcommand\abstractname{Abstract}

\newcommand{\nomenclature}
              [1]
              {
               \bgroup
               \flushleft
               \small\bf
               #1
               \par
               \egroup
              }

\renewcommand{\section}
              [1]
              {
               \bgroup
               \flushleft
               \small\bf
               \refstepcounter{section}
               \arabic{section}. #1
               \par
               \egroup
              }

\renewcommand{\subsection}
              [1]
              {
               \bgroup
               \flushleft
               \small\em
               \refstepcounter{subsection}
               \arabic{section}.
               \arabic{subsection}. #1
               \par
               \egroup
              }

\renewcommand{\subsubsection}
              [1]
              {
               \bgroup
               \flushleft
               \small\em
               \refstepcounter{subsubsection}
               \arabic{section}.
               \arabic{subsection}.
               \arabic{subsubsection}. #1
               \par
               \egroup
              }

  \newcommand{\acknowledgement}
              [1]
              {
               \bgroup
               \flushleft
               \small\bf
               #1
               \par
               \egroup
              }

  \newcommand{\sectionbib}
              [1]
              {
               \bgroup
               \flushleft
               \small\bf
               #1
               \par
               \egroup
              }

\begin{document}



\small
\baselineskip 10pt

\setcounter{page}{1}
\title{\LARGE \bf Direct Numerical Simulation of MILD Combustion: Mixing and Autoignition from Non-Premixed Streams}

\author{{\large Lorenzo Frascino$^{a,*}$, Gandolfo Scialabba$^{a}$, Hongchao Chu$^{a}$, Heinz Pitsch$^{a}$}\\[10pt]
        {\footnotesize \em $^a$Institute for Combustion Technology, RWTH Aachen University, Aachen, Germany }\\[-5pt]}

\date{}  

\twocolumn[\begin{@twocolumnfalse}
\maketitle
\rule{\textwidth}{0.5pt}
\vspace{-5pt}

\begin{abstract} 
Moderate or intense low-oxygen dilution (MILD) combustion is achieved by strongly diluting and preheating the reactants through mixing with hot combustion products before ignition. To better understand how fuel/air/product mixing and interaction govern MILD combustion dynamics, a novel direct numerical simulation (DNS) dataset of a temporally evolving three-stream mixing layer consisting of fuel, air, and hot combustion products has been performed. In this configuration, both fuel–air and air–hot products mixing processes are considered with varying time scales, through four carefully designed DNS cases, to assess how their combined interaction controls ignition under MILD conditions. It is observed that the cases with higher dilution levels fall within the MILD combustion regime, whereas those with lower dilution correspond to non-MILD conditions. The results show that, as long as MILD conditions are observed, ignition is mainly driven by mixing with hot products. Flame index (FI) and local equivalence ratio ($\phi$), combined with chemical explosive mode analysis (CEMA), further identifies the local combustion mode: in MILD cases, ignition occurs predominantly through a premixed-autoignition mode, while in non-MILD scenarios, the premixed-deflagrative contribution to the heat release rate is more substantial. Conditional analysis of scalar dissipation rates shows that the combustion modes in MILD conditions are sensitive to mixing by both the fuel and hot products, whereas the combustion modes in non-MILD conditions are mainly influenced by the mixing of the fuel with the surrounding gases. These findings indicate that MILD combustion exhibits predominantly autoignition-dominated dynamics that remain strongly coupled with multi-stream mixing, while non-MILD regimes remain characterized by stratified, flame-propagation–dominated dynamics.
\end{abstract}

\vspace{10pt}

{\bf Novelty and significance statement}

\vspace{10pt}
This work introduces a new DNS dataset specifically designed to investigate turbulent MILD combustion under practically relevant non-premixed inlet conditions, and, unlike previous studies, explicitly considers the coupled mixing of fuel, air, and hot products. The results demonstrate that, across all combinations of fast and slow mixing for the different streams, as long as MILD conditions are satisfied, the case exhibits premixed-autoignition-dominated characteristics. This highlights the ratio between hot products mixing time and minimum ignition delay as the governing mechanism distinguishing MILD from non-MILD behavior. The study, combining flame index and CEMA, also reveals negligible deflagrative and diffusive heat-release contributions. These datasets clarify how different mixing scenarios influence the establishment of MILD conditions, providing a practical guideline for MILD system design. Its availability is expected to support the community in the validation and development of reduced-order combustion models and to inform the design of systems operating in MILD regimes.

\vspace{5pt}
\parbox{1.0\textwidth}{\footnotesize {\em Keywords:} MILD combustion; Direct numerical simulation; CEMA;}
\rule{\textwidth}{0.5pt}
*Corresponding author.
\vspace{5pt}
\end{@twocolumnfalse}] 

\section{Introduction\label{sec:introduction}} \addvspace{10pt}

Moderate or intense low-oxygen dilution (MILD) combustion enables stable, low-emission operation, for instance, in industrial high-temperature process heating, by promoting strong mixing of hot combustion products with fuel and air prior to ignition~\cite{Cavaliere2004}. This leads to preheated and diluted reactants and a moderate post-ignition temperature rise~\cite{Wunning1997}.
According to the Cavaliere-de Joannon criterion~\cite{Cavaliere2004}, MILD conditions are achieved when the inlet temperature $T_\mathrm{in}$ exceeds the self-ignition temperature $T_\mathrm{si}$ of the reactant mixture, while the maximum temperature rise during the combustion process remains below $T_\mathrm{si}$, i.e., $T_\mathrm{max}-T_\mathrm{in}<T_\mathrm{si}$. Although no absolute temperature threshold is imposed by this definition, a practical upper bound of $T_\mathrm{max} < \SI{1800}{K}$ is commonly adopted in the literature~\cite{deJoannon2000, Iavarone2020} to ensure suppression of thermal NO$_\mathrm{x}$ formation, which becomes significant above this temperature through the Zeldovich mechanism~\cite{Zeldovich1946}.
From the modeling perspective, the absence of a distinct reactive layer and the occurrence of multiple ignition events distributed in space and time further complicate the description of scalar dissipation, progress variable definition, and reaction closure, putting the application of conventional reduced-order models for such conditions in question~\cite{Sorrentino2024}. A detailed understanding of how mixing intensity and stratification influence ignition and heat release in MILD combustion is therefore essential for model development and validation. 

Previous direct numerical simulation (DNS) studies~\cite{Minamoto2013, Minamoto2014, Swaminathan2019} have provided valuable insights into distributed reaction zones and flame--flame interactions under MILD conditions. The structure of reaction zones in non-premixed MILD combustion with internal exhaust gas recirculation was further investigated by Doan et al.~\cite{Doan2018, Doan2019}, who identified the coexistence of ignition fronts and propagating flames, and showed that the dominant combustion mode depends strongly on the mixture fraction lengthscale. 
Subsequently, Doan et al.~\cite{Doan2021} applied chemical explosive mode analysis (CEMA) to both premixed and non-premixed MILD combustion DNS in a freely decaying homogeneous isotropic turbulence (HIT) configuration, showing that premixed MILD flames behave predominantly as autoignition waves, while non-premixed cases exhibit a varying balance depending on mixture-fraction stratification. A step toward more realistic configurations was taken by van Oijen~\cite{vanOijen2013}, and Goktolga et al.~\cite{vanOijen2015}, who investigated a jet-in-hot-coflow (JHC) setup~\cite{Dally2002} through DNS of a temporally evolving mixing layer between lean combustion products and fuel. These studies provided valuable insight into preferential diffusion effects and the transition between flame propagation and autoignition under diluted conditions. 

However, existing DNS studies do not explicitly consider the coupled mixing of fuel, air, and hot combustion products, which is critical to represent the interactions between reacting and non-reacting streams in practical non-premixed environments. As a result, the influence of mixing intensity and scalar stratification on local combustion regimes under non-premixed MILD conditions has not yet been systematically analyzed. In this work, this is investigated using DNS of a temporally evolving mixing layer that involves three streams of fuel, air, and hot combustion products, considering their mixing process before ignition. A systematic parametric study is conducted by independently varying the fuel–air and hot products–air mixing to quantify their impact on the combustion regime evolution, with a focus on the balance between autoignition and deflagration modes. The analysis employs and combines the flame index (FI) and the local equivalence ratio ($\phi$) with CEMA to characterize local combustion modes and provide both physical and modeling-relevant insights. The paper is structured as follows: Section~\ref{sec:setup} defines the DNS configuration and its relevant parameters. In Section~\ref{sec:results}, an analysis of the influence of mixing intensity on ignition behavior is provided. CEMA is then employed to characterize the distribution of combustion modes and, in combination with FI and $\phi$, to characterize flame propagation, thereby discussing the associated modeling implications. Section~\ref{sec:conclusions} summarizes the main findings.

\section{Key parameters and simulation
setup\label{sec:setup}} \addvspace{10pt}

\subsection{DNS configuration\label{subsec:subsection}} \addvspace{10pt}

\begin{figure}[h!]
\centering
\includegraphics[width=192pt]{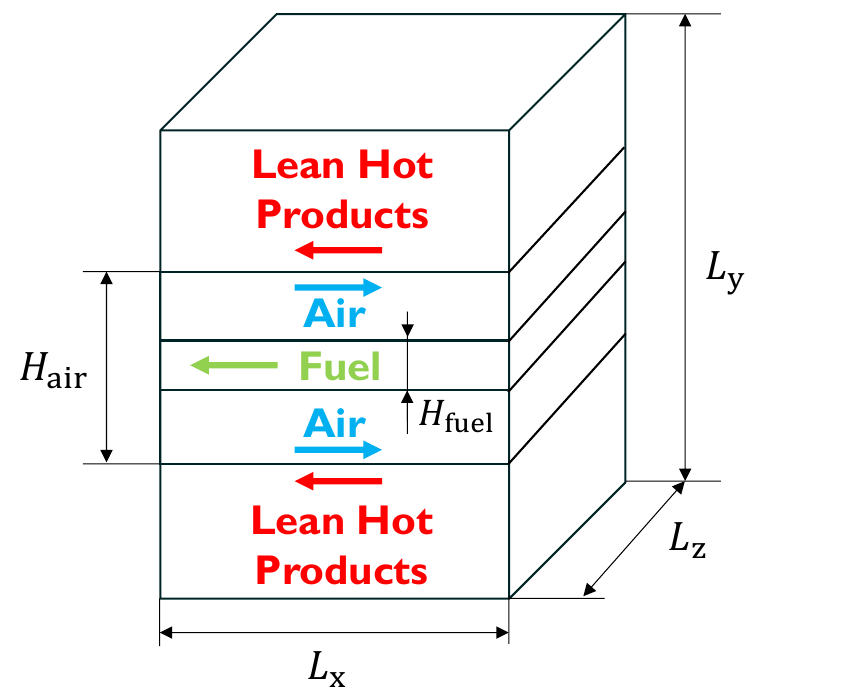}
\caption{\footnotesize Schematic of the numerical configuration.}
\label{sketch}
\end{figure}

To mimic the local mixing conditions and combustion physics of a reverse-flow MILD combustion furnace~\cite{Arghode2011reverse, Ferrarotti2020}, a DNS of a temporally evolving mixing layer is performed. Such a configuration allows the analysis of longer physical mixing phenomena in a compact domain~\cite{Hawkes2012, Attili2014, Scialabba2025}, thereby overcoming the prohibitive computational costs of spatially evolving configurations. A schematic of the performed DNS cases is provided in Fig.~\ref{sketch}. To better understand the relative mixing contributions, fuel, air, and hot products are initialized as separate streams, as depicted in the schematic.
This configuration enables the independent variation of fuel–air and hot products–air mixing time scales, thus controlling the corresponding Damköhler numbers and allowing a systematic assessment of their impact on ignition behavior and MILD combustion establishment.
The numerical domain consists of a box with periodic boundary conditions in the streamwise (x) and crosswise (z) directions, while an outlet boundary condition is imposed in the spanwise (y) direction. The central cold fuel jet composition is $25\%$-vol hydrogen and $75\%$-vol methane at $T_{\mathrm{fuel}} = \SI{300}{K}$.
Two jets of preheated air at $T_{\mathrm{air}} = \SI{900}{K}$ are placed on the sides of the fuel jet, with an opposing velocity direction to favor their relative mixing.
Finally, the air jets are surrounded by two additional jets composed of equilibrium combustion products at $T_{\mathrm{hot}} = \SI{1225}{K}$.
For the selected case, fuel and air correspond to a volumetric global equivalence ratio $ER$ of 0.8; this requirement constrains the relative size of the two jets. The same equivalence ratio was also used to define the composition of the combustion products in the hot stream. 
The jet direction is defined by the velocity imposed on the two shear-layer regions.
The initial conditions for the DNS (temperature, composition, velocity range, and operating pressure) were extracted from a reference experimental campaign by Ferrarotti et al.~\cite{Ferrarotti2020}. The chosen case corresponds to one of the operating conditions identified in the experimental campaign as representative of MILD combustion conditions.
The choice of opposing velocity directions for the streams follows the canonical temporal mixing layer formulation, in which the two streams are initialized symmetrically at $\pm\Delta U/2$~\cite{RogersMoser1994, Almagro2017}. This configuration is equivalent to the experimental setup of~\cite{Ferrarotti2020} in the convective reference frame: for a temporal mixing layer, the only dynamically relevant parameter is the velocity difference $\Delta U$ between the streams, not their absolute velocities~\cite{RogersMoser1994}.

The initial fields for temperature and species mass fractions are mapped using two separate non-premixed 1D flamelet simulations at extinction conditions, each representing one of the mixing systems (fuel-air, air-hot products). 
Using extinct flamelets ensures that the thermochemical initial condition is in a quasi-frozen state, allowing the simulation to capture the onset of autoignition. 
The mapping is done on a smooth mixture fraction profile obtained from a piecewise definition with linear variation across the transition layer.
The 1D flamelets were computed with FlameMaster~\cite{FLAMEMASTER} using a reduced mechanism for lean methane-hydrogen blend combustion with 24 species and 251 reactions, which was derived from the full mechanism C3MechV3.3 model developed by Dong et al.~\cite {Dong2022}. The same kinetic mechanism is employed for the 3D simulation. For the DNS, the reactive, unsteady Navier-Stokes equations are solved in the low-Mach limit using the in-house finite-differences solver CIAO~\cite{Desjardins2008}. Mass diffusion coefficients are computed using a mixture-averaged transport model to account for non-unity Lewis number effects.
\subsection{Parameter definitions\label{subsec:subsection}} \addvspace{10pt}

\begin{figure}[h!]
\centering
\includegraphics[width=192pt]{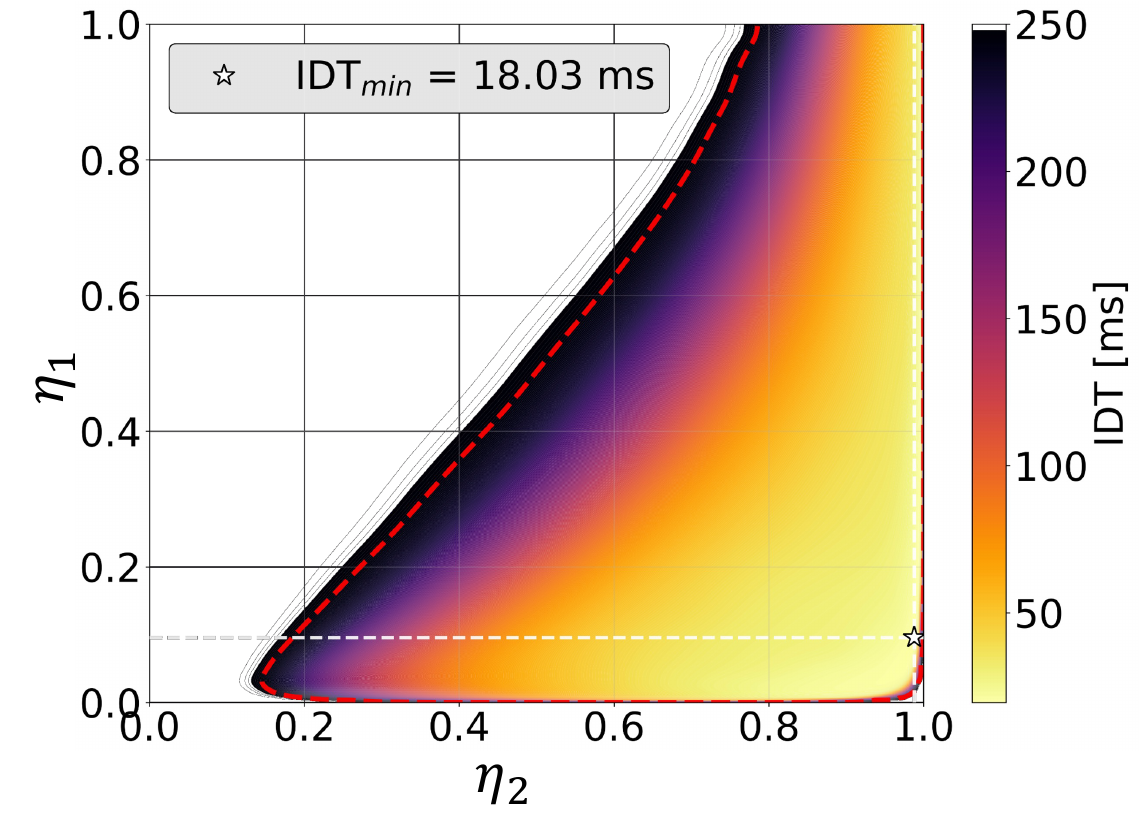}
\caption{\footnotesize Two-dimensional ignition delay time (IDT) map as a function of the parametric coordinates $\eta_1$ and $\eta_2$. The red dashed line denotes the ignition boundary, defined as the locus of compositions whose IDT equals the simulation time limit of 250~ms. All points beyond this boundary are assigned a nominal IDT value of 250~ms for visualization purposes. The white dashed lines indicate the coordinates of the minimum ignition delay time, aiding its identification in the mixture fraction space.}
\label{fig:IDT}
\end{figure}

The fuel/air/products system is described as a ternary mixture using the following mixture fractions:
\begin{itemize}
\item $Z_{\mathrm{hot}}$: mixture fraction associated with the hot products stream. $Z_{\mathrm{hot}} = 1$ corresponds to pure hot products, while $Z_{\mathrm{hot}} = 0$ indicates the absence of hot products, i.e., a mixture of fuel and/or air.

\item $Z_{\mathrm{fuel}}$: mixture fraction associated with the fuel stream. $Z_{\mathrm{fuel}} = 1$ corresponds to pure fuel, while $Z_{\mathrm{fuel}} = 0$ indicates the absence of fuel, i.e., a mixture of air and/or hot products.

\item $Z_{\mathrm{air}} = 1 - Z_{\mathrm{fuel}} - Z_{\mathrm{hot}}$: complementary air mixture fraction.
\end{itemize}
The process is characterized by two mixing dynamics: one between fuel and air (FA) and another between hot products and air (HA). These two processes are described using two Reynolds numbers defined as 
\begin{equation}
Re_{\mathrm{FA}} = \frac{(H_{\mathrm{fuel}} \, \Delta U_{\mathrm{FA}})}{\nu_{\mathrm{fuel}}}\,, \quad
Re_{\mathrm{HA}} = \frac{(H_{\mathrm{air}} \, \Delta U_{\mathrm{HA}})}{\nu_{\mathrm{air}}}\,,
\end{equation}
where $\Delta U_{\mathrm{FA}}$ and $\Delta U_{\mathrm{HA}}$ are the bulk velocity differences between the fuel and air jets and between the hot products and air jets. $H$ is the characteristic jet width. For the hot products-air mixing dynamics, the reference width $H_{\mathrm{air}}$ is taken as the sum of the widths of the two air jets and the central fuel jet; $\nu$ is the kinematic viscosity for either fuel or air. The mixing time scales are defined as
\begin{equation}
\tau_{\mathrm{FA}} = \frac{4.5 H_{\mathrm{fuel}}}{\Delta U_{\mathrm{FA}}}\,, \quad \tau_{\mathrm{HA}} = \frac{4.5 H_{\mathrm{air}}}{\Delta U_{\mathrm{HA}}}\,,
\end{equation}
following Pope’s formulation~\cite{POPE}. The Damköhler numbers are the ratios between these mixing times and a representative chemical time $\tau_{\mathrm{chem}}$, obtained from 0D homogeneous reactor simulations, performed with FlameMaster~\cite{FLAMEMASTER}, over the 2D $(Z_\mathrm{fuel}, Z_\mathrm{hot})$ domain. This allows the 0D simulations to cover all the possible mixing conditions between the three streams. For each local composition, the ignition delay time was defined as the time required for the temperature to rise by $\SI{10}{K}$ above its initial value~\cite{Sabia2013}. The resulting 2D ignition delay time (IDT) map is shown in Fig.~\ref{fig:IDT}, where the axes represent the parametric coordinates $\eta_1$ and $\eta_2$, with $\eta_1 = Z_{\mathrm{fuel}}$ and $\eta_2 = Z_{\mathrm{hot}}/(1-Z_{\mathrm{fuel}})$. This parametrization ensures that all compositions in the $[0,1]^2$ domain are physically admissible by construction. The red dashed line indicates the ignition boundary within the 0D reactor simulation time limit of 250~ms (which is also the largest physical time reached among all the performed DNS).
The map shows a strong dependence of the ignition delay time on the relative mixing between the interacting streams. The minimum IDT computed is $\tau_{\mathrm{chem}} = \SI{18}{ms}$, and is taken as the characteristic chemical time scale. It represents a fundamental lower bound for ignition delay, since earlier ignition is constrained by diffusive losses~\cite{Mastorakos2009}. Four simulation cases are considered to assess the effects of dilution and fuel–air mixing on ignition behavior. The cases are defined by varying the characteristic Damköhler numbers associated with the hot products-air and fuel–air mixing processes, denoted as $Da_{\mathrm{HA}}$ and $Da_{\mathrm{FA}}$, respectively.
Low $Da_{\mathrm{HA}}$ cases correspond to high-dilution (HD) conditions, while high $Da_{\mathrm{HA}}$ values are associated with low-dilution (LD) regimes.
To further investigate the influence of fuel–air mixing, each dilution condition is combined with two different fuel–air mixing rates: fast fuel mixing (FF) for low $Da_{\mathrm{FA}}$ and slow fuel mixing (SF) for high $Da_{\mathrm{FA}}$. This results in four cases in total: HD-FF, HD-SF, LD-FF, and LD-SF. Between the LD and HD cases, the domain size has been rescaled to maintain a constant Reynolds number of the HA shear layer system across all cases. This approach allows the Damköhler number to be varied as desired without changing the chemical time scale, thereby affecting only the mixing time. The minimum Kolmogorov scale, defined as $\bar{\eta}_k = \bar{\nu}^{3/4} \bar{\varepsilon}^{-1/4}$ and computed as in Ref.~\cite{Pantano2003}, averaging along the $x$ and $z$ directions, is larger than half of the grid size ($\eta_k > \Delta/2$) at all locations and times. In addition, the computational grid in the spanwise direction $y$ is generated using a hyperbolic-sine stretching function, ensuring a smooth and monotone distribution of grid points with enhanced resolution near the fuel jet centerline.
The computational grid consists of approximately 1.2 billion cells for the HD cases and 0.7 billion cells for the LD cases, with a minimum resolution of 10 grid points across the OH layers. 
A summary of the key parameters for the four DNS cases is reported in Table \ref{tab:dns_single}. Further details on the initialization and the scalar mapping procedure are provided in the supplementary material.
\begin{table}[h!] \centering
\caption{DNS parameters for the four considered cases.}
\label{tab:dns_single}
\resizebox{\columnwidth}{!}{
\begin{tabular}{lcccc}
\toprule
Case Name & \textbf{HD-FF} & \textbf{HD-SF} & \textbf{LD-FF} & \textbf{LD-SF} \\
\midrule
$X_{\mathrm{H_2}}$ [\%]  & 25 & 25 & 25 & 25 \\
$X_{\mathrm{CH_4}}$ [\%] & 75 & 75 & 75 & 75 \\
$ER$       & 0.8 & 0.8 & 0.8 & 0.8 \\
$T_{\mathrm{hot}}$ [K]   & 1225 & 1225 & 1225 & 1225 \\
$H_{\mathrm{fuel}}$ [mm] & 0.8 & 0.8 & 3.9 & 3.9 \\
$H_{\mathrm{air}}$ [mm]  & 25 & 25 & 124 & 124 \\
$N_{\mathrm{points}}$ [$10^9$] & 1.2 & 1.2 & 0.7 & 0.7 \\
$L_\mathrm{x}/H{_\mathrm{air}}$  & 30 & 30 & 15 & 15 \\
$L_\mathrm{y}/H{_\mathrm{air}}$  & 30 & 30 & 15 & 15 \\
$L_\mathrm{z}/H{_\mathrm{air}}$  & 15 & 15 & 7.5 & 7.5 \\
$\Delta U_{\mathrm{FA}}$ [m/s] & 1.30 & 0.05 & 6.52 & 0.25 \\
$\Delta U_{\mathrm{HA}}$ [m/s] & 40 & 40 & 8.03 & 8.03 \\
$Re_{\mathrm{FA}}$       & 100 & 2 & 1129 & 43 \\
$Re_{\mathrm{HA}}$       & 10000 & 10000 & 10000 & 10000 \\
$Da_{\mathrm{FA}}$       & 0.2 & 5 & 0.2 & 5 \\
$Da_{\mathrm{HA}}$       & 0.2 & 0.2 & 5 & 5 \\
Total time [ms]          & 60 & 60 & 170 & 250 \\
\bottomrule
\end{tabular}
}
\end{table}

\section{Results and discussion\label{sec:results}} \addvspace{10pt}

\subsection{Ignition evolution and mixing dynamics analysis \label{subsec:temperature_description}}\addvspace{10pt}

\begin{figure*}[h!]
\centering
\vspace{-0.4 in}
\includegraphics[width=\textwidth]{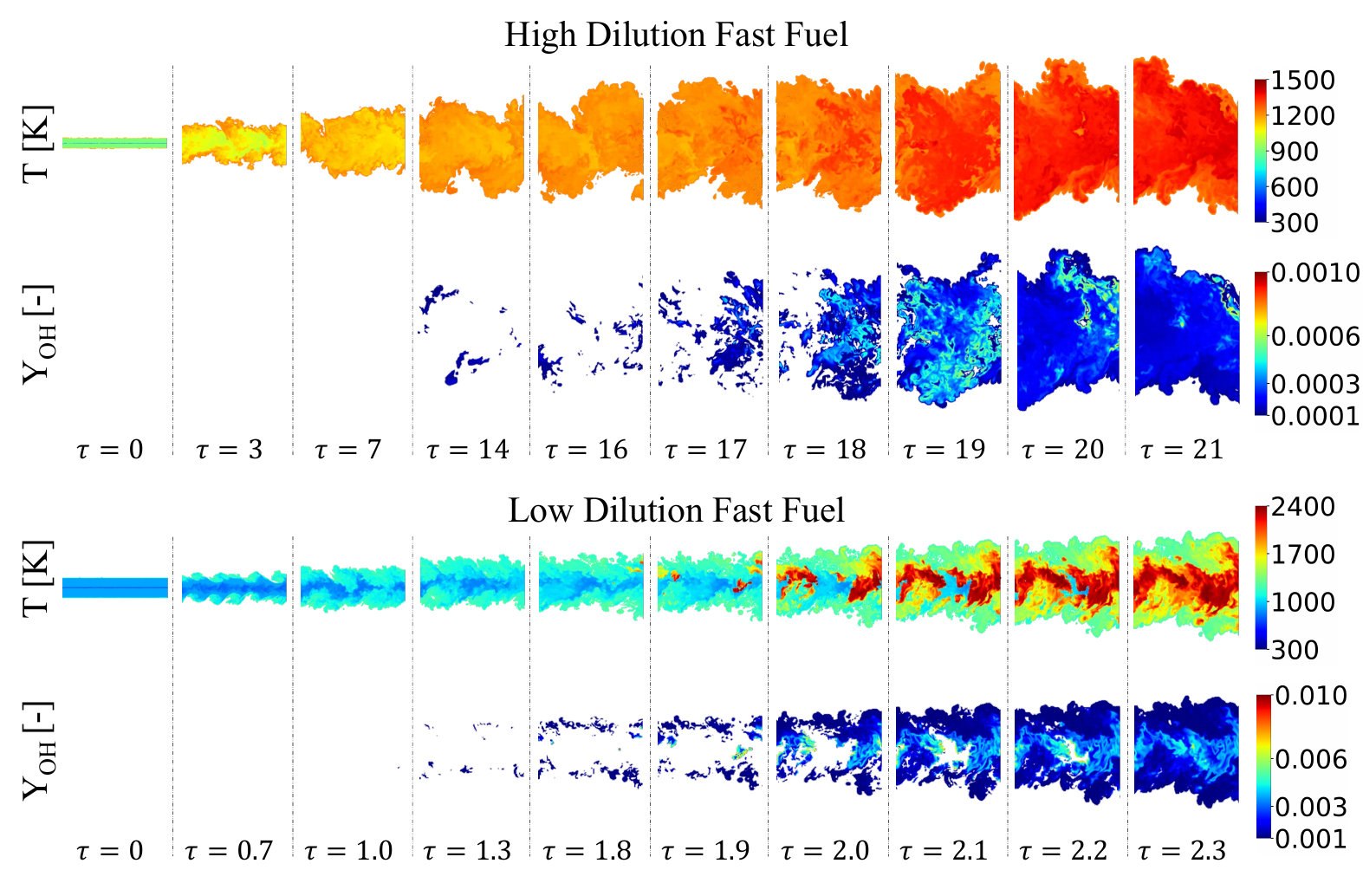}
\caption{Two-dimensional temperature and OH mass fraction field slices for high-dilution (top) and low-dilution (bottom) fast-fuel mixing cases at representative time instants. The reference time $\tau$ shown below each slice corresponds to the ratio between the physical time $t$ and the respective case mixing time ($\tau_{\mathrm{HA}}$).}
\label{HDFF_slices}
\end{figure*}

The results from the DNS study are presented below, highlighting the effects of dilution level and fuel–air mixing intensity on ignition behavior and flame structure. To provide a clear visual understanding of the differences between high-dilution (HD) and low-dilution (LD) cases, Fig.~\ref{HDFF_slices} shows two-dimensional temperature and OH mass fraction slices for the two cases with fast fuel mixing (FF).
The reference time $\tau$ shown below each slice corresponds to the ratio between the physical time and the respective mixing time ($\tau_{\mathrm{HA}}$) of each case.
In the high-dilution case, the intense mixing of hot products leads to reduced temperature peaks and a progressive, spatially distributed ignition process. The $T_{\mathrm{max}}$ reached is \SI{1415}{K}, with $T_\mathrm{in} = \SI{900}{K}$ and a self-ignition temperature around \SI{866}{K}, thus satisfying the Cavaliere-de Joannon criterion for MILD conditions. The temporal sequence shows a gradual heating of the jet and the formation of broad reactive regions, with OH radicals distributed throughout the domain rather than confined in thin reaction layers, characteristic of MILD combustion. In contrast, the low-dilution case exhibits significantly higher peak temperatures, reaching about \SI{2470}{K}, thus, well above the temperature limit for MILD conditions. The reacting layers remain spatially confined and distinct, with sharper gradients and localized regions of high OH concentration typical of conventional turbulent flames. Increasing dilution transforms the flame topology from spatially confined to volumetric combustion, marking the transition from conventional (non-MILD) to MILD regimes. The complete set of instantaneous temperature and OH fields for the four cases is provided in the supplementary material.

Fig.~\ref{Tmean_vs_time} shows the temporal evolution of the mean temperature and the coefficient of determination ($R^2$) between $Z_{\mathrm{air}}$ and $Z_{\mathrm{fuel}}$, addressing the impact of variations in FA mixing intensity.
The mean is computed using all points for which the IDT of the corresponding local mixture is below the maximum simulation time (250~ms), i.e., all points for which the mixture fraction values are within the dashed lines in Fig.~\ref{fig:IDT}.
The vertical dashed lines mark the IDT of each DNS, defined as the instant when the mean temperature exceeds the hot products stream temperature ($T_{\mathrm{hot}} = \SI{1225}{K}$).
\begin{figure}[h!]
\centering
\includegraphics[width=192pt]{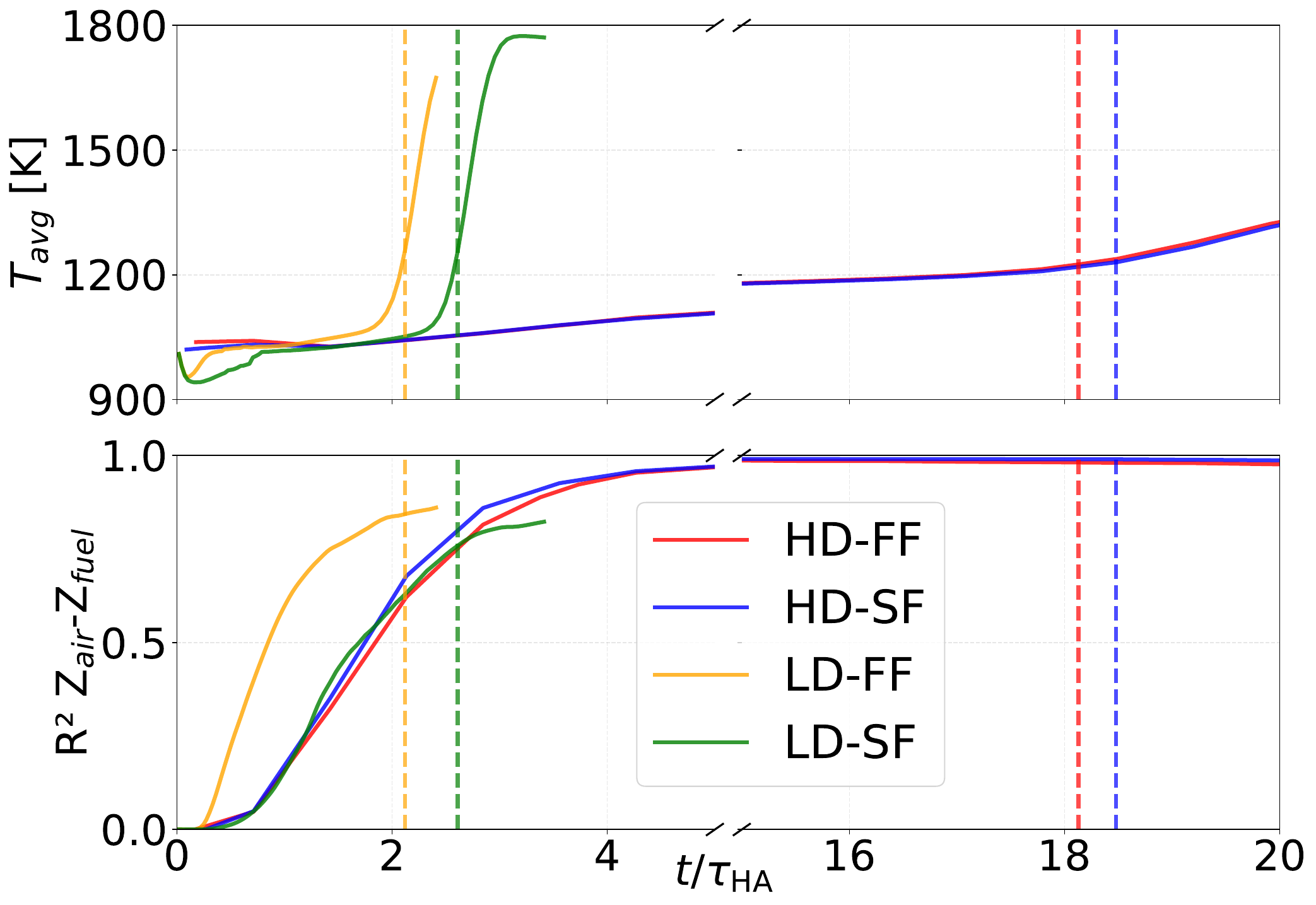}
\caption{\footnotesize (Top) mean temperature evolution computed over the regions of the 2D ignition map (Fig~\ref{fig:IDT}) that ignited within the simulated physical time equal to $\SI{250}{ms}$ and (bottom) temporal evolution of the coefficient of determination ($R^2$) between $Z_{\mathrm{air}}$ and $Z_{\mathrm{fuel}}$. The ignition times are indicated with dashed vertical lines, color-coded consistently with the corresponding curves. The time axis has been non-dimensionalized by the respective mixing time $\tau_{\mathrm{HA}}$ of each case.}
\label{Tmean_vs_time}
\end{figure}
Both LD cases show higher temperatures, and ignition occurs later for the LD-SF case, suggesting that fuel–air mixing affects ignition when MILD conditions are not met.
Interestingly, both the FF- and SF-HD cases exhibit low mean temperatures, with nearly identical profiles and only minor differences in IDT. This indicates that the fuel–air mixing intensity ($Da_{\mathrm{FA}}$) has a limited impact once MILD conditions are established. However, this behavior may be influenced by both the small fuel-to-air mass ratio and the stronger turbulence associated with the hot products-air stream, which promotes mixing. The $R^2$ evolution further confirms this behavior: under non-MILD conditions, ignition occurs while $R^2$ remains well below unity, indicating partial mixing and local stratification; for MILD cases, $R^2$ rapidly approaches 0.99 before ignition, revealing nearly perfect correlation between fuel and air mixture fractions. These results confirm that the ratio between the mixing time of the hot products shear layer system and the computed minimum ignition delay is a key parameter for establishing MILD conditions.

\subsection{Flame propagation mode characterization\label{subsec:flameindex}} \addvspace{10pt}

\begin{figure*}[h!]
\centering
\vspace{-0.4 in}
\includegraphics[width=\textwidth]{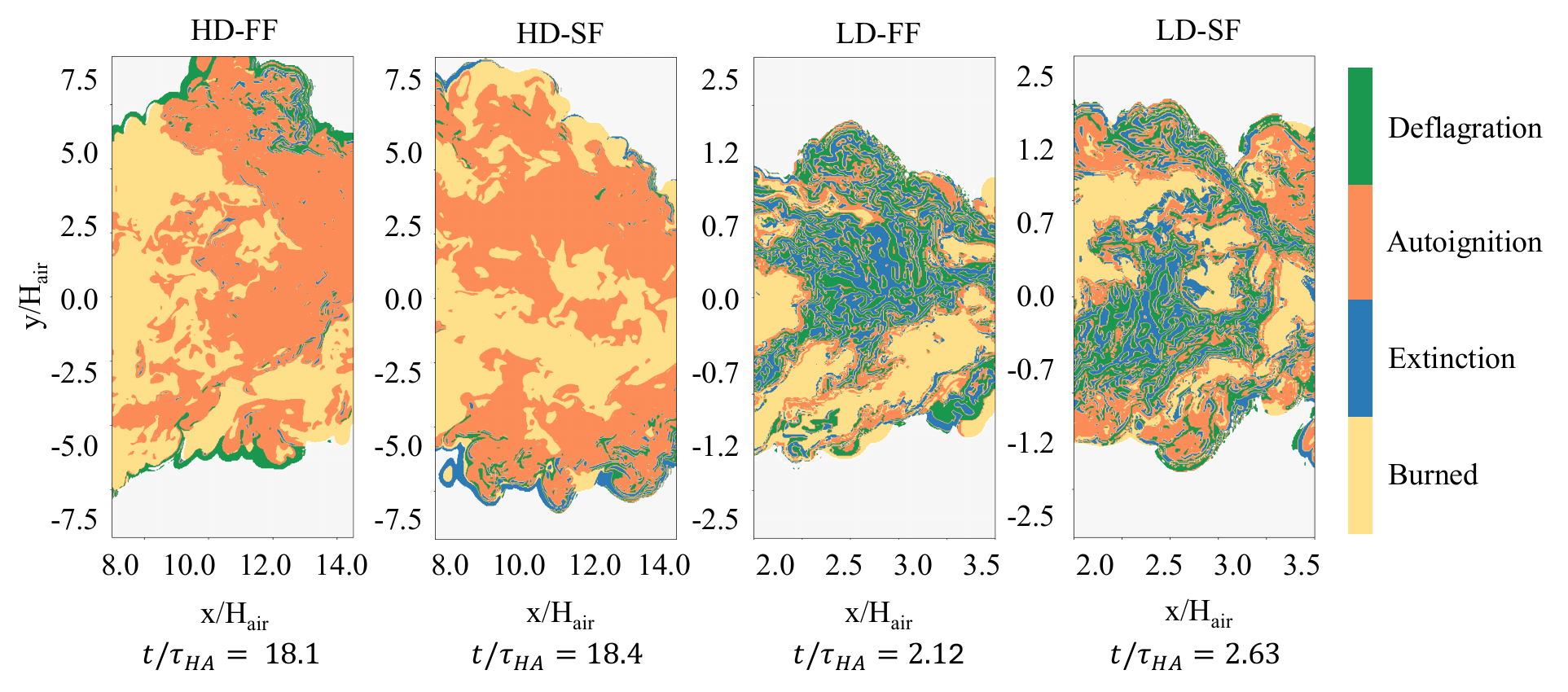}
\caption{Two-dimensional slices of the local mode indicator $\alpha$ obtained from CEMA where the different colors indicate distinct flame propagation modes: orange highlights regions dominated by chemical autoignition onset, green corresponds to stable flame propagation (deflagration), yellow represents non–reactive or weakly reactive mixtures (post-ignition zones), and blue identifies areas where suppression of reactivity occurs. For each case, the data around the ignition time defined in the previous section is used.}
\label{CEMA_CUTS}
\end{figure*}
\begin{figure}[h!]
\centering
\includegraphics[width=0.48\textwidth]{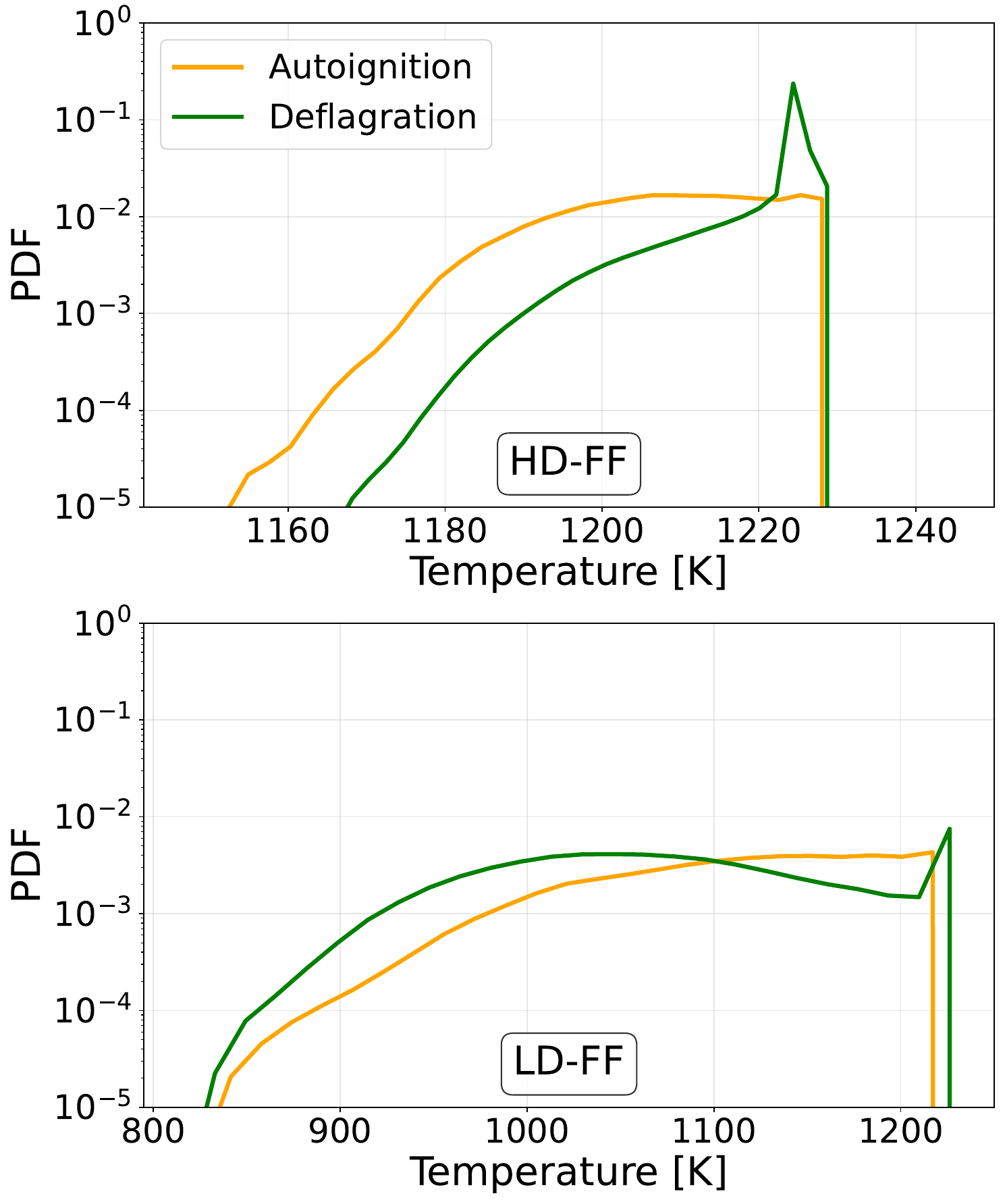}
\caption{\footnotesize Marginal PDFs of the temperature for a given combustion mode (autoignition and deflagration) for the HD-FF (top) and LD-FF (bottom) cases, computed in the pre-ignition regions at the ignition time $\tau_{\mathrm{ign}}$ of each respective case.}
\label{fig:PDF_TEMP_CEMA}
\end{figure}
\begin{figure}[h!]
\centering
\includegraphics[width=0.48\textwidth]{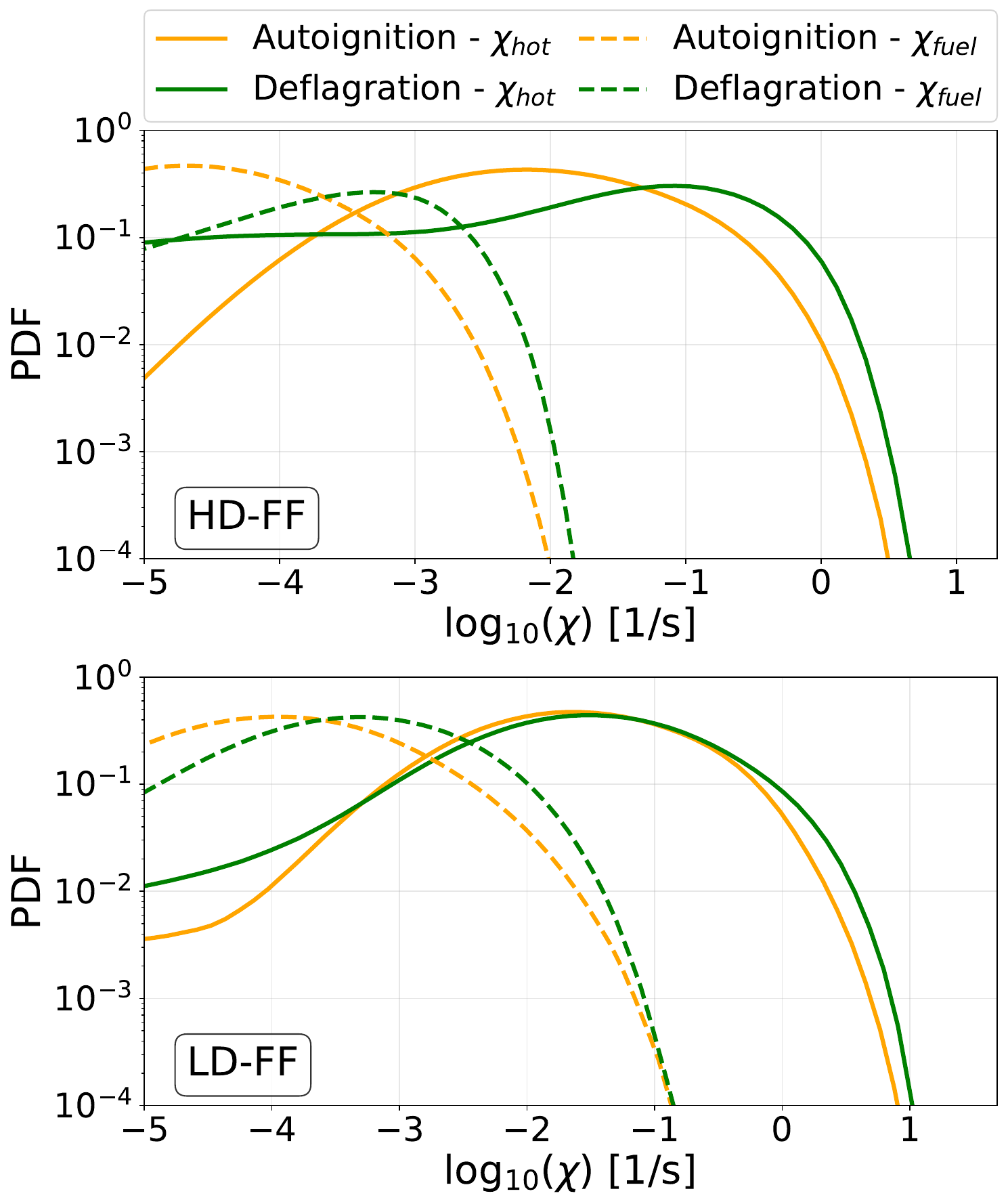}
\caption{Marginal PDFs of scalar dissipation rates $\chi_{\mathrm{hot}}$ (solid lines) and $\chi_{\mathrm{fuel}}$ (dashed lines) conditioned on combustion mode (autoignition and deflagration) for HD-FF (top) and LD-FF (bottom) cases in pre-ignition regions at the ignition time $\tau_{\mathrm{ign}}$ of each respective case.}
\label{fig:PDF_CHI_CEMA}
\end{figure}

To further characterize the local combustion dynamics, the chemical explosive mode analysis (CEMA) is employed. With such analysis, it can be defined whether the system evolves under autoignition conditions (0D-reactor-like behavior) or, instead, it exhibits deflagration-like behavior. 
Originally introduced by Lu et al.~\cite{CEMA_1} and later extended to include diffusion effects by Xu et al.~\cite{CEMA_2}, CEMA is based on the analysis of the eigenvalues $\lambda_\mathrm{e}$ of the Jacobian matrix $J_\mathrm{\omega}$ of the chemical source term $\omega$.
It is noted that CEMA presents some limitations with respect to more complete frameworks such as Computational Singular Perturbation~\cite{Lam1989,Goussis2021}; a comparative assessment against the latter is reported in the supplementary material, where it is shown that both methods yield identical classifications in the pre- and early-ignition region of interest. 
In addition, CEMA is adopted here to enable a direct qualitative comparison with the DNS study of Doan et al.~\cite{Doan2021}, who employed the same methodology in a non-premixed MILD combustion configuration.
A local mode indicator is introduced as
\begin{equation}
\alpha = \frac{\phi_\mathrm{s}}{\phi_{\mathrm{\omega}}}\,,
\end{equation}
which quantifies the relative role of diffusion and chemistry, with $\phi_{\mathrm{\omega}}$ representing the chemical contribution and $\phi_{\mathrm{s}}$ the non-chemical one. Further details on the method and the computation of $\phi_\mathbf{s}$ and $\phi_\mathrm{\omega}$ can be found in~\cite{CEMA_2} and in the supplementary material. The value of $\alpha$ identifies different combustion modes:  
\begin{itemize}
    \item $\alpha > 1$: assisted-ignition mode, dominated by diffusion (deflagration-like behavior);  
    \item $|\alpha| < 1$: autoignition mode, governed primarily by chemistry;  
    \item $\alpha < -1$: extinguishing mode, where diffusion suppresses chemistry.  
\end{itemize}
Fig.~\ref{CEMA_CUTS} presents 2D slices of the local mode indicator for the four DNS cases, evaluated at their respective characteristic ignition instants. 
From the $\alpha$ fields, no significant change is observed when varying the fuel-air mixing intensity (i.e., HD cases): reactive structures and $\alpha$ values remain very similar across cases with the same dilution level. The main difference, however, emerges when comparing MILD (HD) and non-MILD (LD) cases. Specifically, under MILD conditions, the main jet core is largely dominated by regions prone to autoignition. In contrast, in non-MILD conditions, the jet core exhibits a wider distribution of regions with deflagration and extinguishing behavior.
Fig.~\ref{fig:PDF_TEMP_CEMA} shows the temperature probability density functions (PDFs) of the autoignition- and deflagration-dominated regions, evaluated in non-ignited zones. In the non-MILD case (LD-FF), the temperature distribution is broad, reflecting significant thermal and compositional stratification in the pre-ignition zones. The presence of colder, less reactive pockets with longer ignition delay times increases the likelihood of flame-propagation events, resulting in a noticeable contribution from deflagration-dominated regions. In contrast, the MILD case (HD-FF) exhibits narrower and more sharply peaked PDFs, indicative of enhanced homogeneity and stronger pre-ignition mixing. This reduced stratification originates from the dilution and preheating effects of the recirculated combustion products, which establish a nearly uniform thermochemical environment.
Despite this, the PDFs reveal that the highest-temperature regions in the pre-ignition zones, which correspond to mixtures in the vicinity of the hot products shear layer (as $T_{\mathrm{hot}} = \SI{1225}{K}$ represents the upper bound of the pre-ignition temperature range), show a higher probability of deflagration-dominated behavior, as also consistent with the $\alpha$ fields shown in Fig.~\ref{CEMA_CUTS}. To investigate this, Fig.~\ref{fig:PDF_CHI_CEMA} shows PDFs of the scalar dissipation rate ($\chi$) in the autoignition and deflagration-dominated zones. This analysis is of particular interest in the context of MILD combustion, as recent work by Sabia et al.~\cite{Sabia2024} has shown that the classical hysteresis behavior with an unstable branch is retained under most MILD combustion conditions, supporting a non-trivial role of scalar dissipation in governing the local combustion mode.
Two dissipation rates can be defined, one associated with mixing of hot products into the other streams ($\chi_{\mathrm{hot}}$) and one associated with mixing of the fuel into other streams ($\chi_{\mathrm{fuel}}$), defined as
\begin{equation}
\chi_{\mathrm{fuel}} = 2D_\mathrm{t} ( \nabla Z_{\mathrm{fuel}} )^2\,,
\quad
\chi_{\mathrm{hot}} = 2D_\mathrm{t} ( \nabla Z_{\mathrm{hot}} )^2\,,
\end{equation}
where $D_\mathrm{t}$ is the thermal diffusivity.
The figure reveals distinct mixing-chemistry interactions depending on the combustion regime. For the LD-FF case (non-MILD), $\chi_{\mathrm{hot}}$ exhibits minimal influence on the combustion mode distribution, as evidenced by the almost overlapping PDF profiles for autoignition and deflagration zones. In contrast, high values of $\chi_{\mathrm{fuel}}$ are predominantly associated with deflagration, indicating that intense fuel mixing disfavors autoignition. In the MILD case (HD-FF), however, both $\chi_{\mathrm{hot}}$ and $\chi_{\mathrm{fuel}}$ exhibit distinct distributions between autoignition and deflagration modes. This indicates that, under MILD conditions, the mixing of hot products influences the local combustion mode along with the fuel mixing. This observed sensitivity to both scalar dissipation rates underscores the importance of hot products recirculation and its interplay with fuel mixing in governing the distributed, autoignition-dominated nature of MILD combustion.

To provide further insight into the flame propagation behavior, the local flame index (FI)~\cite{Flame_index} is additionally used to distinguish between premixed and diffusion-dominated regions, defined as
\begin{equation}
    \text{FI} = \nabla Y_{\mathrm{fuel}} \cdot \nabla Y_{\mathrm{oxidizer}}\,.
\end{equation}
Positive values of FI indicate premixed regions where fuel and oxidizer gradients are aligned, while negative values correspond to diffusion-controlled zones (non-premixed). The integral heat release rate (HRR) was then evaluated by combining CEMA and FI, conditioning both the combustion mode (autoignition, deflagration and extinguishing, identified by $\alpha$) and the flame type (premixed and non-premixed, identified by FI) at the ignition time $\tau_{\mathrm{ign}}$ of each respective case, consistently with the fields shown in Figs.~\ref{CEMA_CUTS},~\ref{fig:PDF_TEMP_CEMA}~and~\ref{fig:PDF_CHI_CEMA}. Additionally, the analysis is further conditioned on the local equivalence ratio $\phi$, defined following Bilger's formulation~\cite{Bilger1990} as
\begin{equation}
\phi = \frac{2 Y_\mathrm{C} / W_\mathrm{C} + 0.5 Y_\mathrm{H} / W_\mathrm{H} - Y_\mathrm{O} / W_\mathrm{O}}
            {(2 Y_\mathrm{C} / W_\mathrm{C} + 0.5 Y_\mathrm{H} / W_\mathrm{H} - Y_\mathrm{O} / W_\mathrm{O})_{\mathrm{st}}}\,,
\end{equation}
where $Y_\mathrm{C}$, $Y_\mathrm{H}$, $Y_\mathrm{O}$ are the elemental mass fractions of carbon, hydrogen, and oxygen, $W_\mathrm{C}$, $W_\mathrm{H}$, $W_\mathrm{O}$ are their atomic masses, and the subscript $\mathrm{st}$ denotes stoichiometric conditions. This allows the HRR contributions to be separately attributed to lean ($\phi < 1$) and rich ($\phi > 1$) mixture regions. 
The fractional HRR contributions obtained for the four cases are shown in Table~\ref{tab:cema_perc}, where A denotes autoignition, D deflagration, P premixed, nP non-premixed, L lean, R rich, and E extinguishing.
\begin{table}[h!] \centering \caption{Spatially integrated heat release rate fractional contributions based on the flame propagation mode, evaluated at the ignition time of each case. The subscripts A, D, P, nP, L, R, E denote autoignition, deflagration, premixed, non-premixed, lean, rich, and extinguishing, respectively. The ignition time $\tau_{\mathrm{ign}}$ is reported as the physical ignition time normalized by the corresponding mixing time $\tau_{\mathrm{HA}}$.}\label{tab:cema_perc} \resizebox{\columnwidth}{!}{ \begin{tabular}{lcccc} \toprule Case Name & \textbf{HD-FF} & \textbf{HD-SF} & \textbf{LD-FF} & \textbf{LD-SF} \\ \midrule 
$HRR_{\mathrm{APL}}$ [\%] & 91.10 & 88.01 & 62.73 & 57.52 \\ 
$HRR_{\mathrm{AnPL}}$ [\%] & 4.04 & 4.21 & 3.29 & 2.69 \\
$HRR_{\mathrm{APR}}$ [\%] & 0.00 & 0.00 & 20.11 & 20.28 \\ 
$HRR_{\mathrm{AnPR}}$ [\%] & 0.00 & 0.00 & 0.92 & 0.69 \\
$HRR_{\mathrm{DPL}}$ [\%] & 2.00 & 4.39 & 4.13 & 5.78 \\ 
$HRR_{\mathrm{DnPL}}$ [\%] & 0.08 & 0.14 & 0.31 & 0.30 \\
$HRR_{\mathrm{DPR}}$ [\%] & 0.00 & 0.00 & 4.38 & 7.05 \\ 
$HRR_{\mathrm{DnPR}}$ [\%] & 0.00 & 0.00 & 0.29 & 0.2 \\ 
$HRR_{\mathrm{E}}$ [\%] & 2.78 & 3.25 & 3.85 & 5.48 \\
$\tau_{\mathrm{ign}} $ & 18.1 & 18.4 & 2.12 & 2.63 \\
\bottomrule \end{tabular} } 
\end{table}
The results highlight clear differences between MILD and non-MILD regimes. Under MILD conditions (HD cases), more than 88\% of the total HRR originates from autoignitive and premixed regions, confirming that combustion proceeds in a volumetric and homogeneous manner, while diffusive and deflagrative contributions remain minor. The nearly identical behavior of HD-FF and HD-SF further confirms that the fuel–air mixing intensity has a limited influence on the flame propagation mode once MILD conditions are established. Notably, under MILD conditions, the HRR contributions from rich mixture regions ($\phi > 1$) are identically zero for both HD cases, indicating that the intense pre-ignition mixing driven by the hot products shear layer produces a thermochemical environment so homogeneous that no distinct rich pockets survive to ignition. This differs from the findings of Doan et al.~\cite{Doan2018}, who observed non-negligible contributions from both lean and rich premixed modes in non-premixed MILD combustion with internal EGR.
In contrast, the non-MILD cases (LD cases) exhibit higher deflagrative activity, with a more significant HRR contribution from diffusion-dominated zones. Furthermore, the non-MILD cases show non-negligible contributions from rich mixture regions, with $HRR_{\mathrm{APR}} \approx 20\%$ and $HRR_{\mathrm{DPR}} \approx 4$--$7\%$, consistent with the higher mixture fraction stratification present in these cases as evidenced by the lower $R^2$ values at ignition. These features correspond to localized reactive layers and non-MILD flame structures. These findings are broadly consistent with those of Doan et al.~\cite{Doan2021}, who similarly observed a dominance of autoignition in non-premixed MILD combustion, with autoignitive regions contributing over 85\% of the total heat release rate. In the present cases, the combined autoignitive contribution reaches approximately 95\% for the HD-FF case and 93\% for the HD-SF case, suggesting a slightly higher prevalence of autoignition than in the non-premixed cases of Doan et al.~\cite{Doan2021}. This difference may be attributed to the shear-driven mixing in the present configuration, which promotes a more uniform thermochemical environment prior to ignition compared to the freely decaying HIT, reducing the probability of flame-propagation events and rich zones. As a result, the HRR is entirely governed by autoignitive regions, demonstrating that, under non-premixed MILD conditions, the system behaves as an autoignition wave rather than a propagating flame.
These findings, together with the distinct scalar dissipation rate sensitivities observed in MILD and non-MILD cases, highlight the need for a flexible modeling framework capable of capturing both combustion regimes. While reduced models based on autoignition chemistry may effectively capture the dominant heat-release dynamics under MILD conditions, the formulation must explicitly account for the coupling between chemical time scales and mixing of both hot products and fuel. In contrast, for non-MILD regimes, the framework must also reproduce flame-propagation dynamics, where the fuel mixing with other gases predominantly governs the autoignition–deflagration competition. The model should thus capture these mixing-chemistry interactions and the transitions between regimes.

\section{Conclusions\label{sec:conclusions}} \addvspace{10pt}
Direct numerical simulations of a temporally evolving, non-premixed mixing layer formed by three interacting fuel, air, and hot combustion products jets have been performed to address the following questions:
(i) How do dilution and fuel–air mixing intensity influence ignition dynamics and flame structure in MILD and non-MILD regimes?
(ii) How do local combustion modes evolve across these conditions, and what are the modeling-relevant insights?

High-dilution cases are observed to exhibit low peak temperatures and spatially distributed radicals, characteristic of MILD combustion. In contrast, lower dilution cases show stronger temperature gradients and stratification, with an increased sensitivity to fuel–air mixing and different ignition delays. This indicates that the relative time scale between hot products mixing and the minimum ignition delay time is the key parameter for establishing MILD conditions, providing a practical guideline for MILD system design. By combining chemical explosive mode analysis (CEMA), flame index (FI), and the local equivalence ratio ($\phi$), it is shown that under MILD conditions, heat release originates almost entirely from lean-premixed-autoignition regions with negligible contribution from diffusive and deflagrative modes. In contrast, non-MILD cases exhibit increasing contributions from deflagrative modes, consistent with the formation of localized reactive layers. The results suggest that the transition to MILD combustion is driven by intense mixing of hot combustion products with reactants, producing a diluted and preheated environment with low thermal and compositional stratification. When ignition occurs, most of the domain reaches autoignition conditions simultaneously, resulting in spatially uniform, autoignition-dominated heat release. This 0D-like behavior indicates that local thermochemical evolution is governed by homogeneous ignition rather than flame propagation. The analysis of scalar dissipation rates shows that, in non-MILD conditions, high values of $\chi_{\mathrm{fuel}}$ are associated with deflagration-dominated combustion, while $\chi_{\mathrm{hot}}$ plays a negligible role. Under MILD conditions, instead, both $\chi_{\mathrm{hot}}$ and $\chi_{\mathrm{fuel}}$ significantly influence the combustion mode, highlighting the importance of mixing of recirculated products in addition to fuel stream mixing.

Overall, these findings highlight the importance of combustion models capable of representing both autoignition- and deflagration-dominated regimes, particularly for configurations where the local combustion mode is not fixed a priori but emerges dynamically from the competition between mixing and chemistry across the different streams. Building on these results, a systematic investigation of the combustion mode balance as a function of different operating conditions, including fuel type, equivalence ratio, hot products temperature, and turbulence intensity, as well as the development of a more general criterion for the ignition-deflagration balance as a function of local flow parameters such as shear rate and mixing layer thickness, represent relevant directions for future work.

\acknowledgement{CRediT authorship contribution statement} \addvspace{10pt}

{\bf Lorenzo Frascino}:  Writing – review \& editing, Writing – original draft, Methodology, Investigation, Formal analysis, Data curation, Conceptualization. {\bf Gandolfo Scialabba}: Writing – review \& editing, Supervision, Methodology, Conceptualization. {\bf Hongchao Chu}:  Writing – review \& editing, Supervision, Methodology, Conceptualization. {\bf Heinz Pitsch}: Writing – review \& editing, Supervision, Project administration, Funding acquisition, Conceptualization.
\acknowledgement{Declaration of competing interest} \addvspace{10pt}

The authors declare that they have no known competing financial interests or personal relationships that could have appeared to influence the work reported in this paper.

\acknowledgement{Acknowledgments} \addvspace{10pt}

L.F. and H.P. acknowledge the funding from the European Union’s Horizon 2020 research and innovation program under the Marie Skłodowska-Curie grant agreement No 101072779. The authors gratefully acknowledge the Gauss Center for Supercomputing e.V. (www.gauss-centre.eu) for funding this project by providing computing time on the GCS Supercomputer Super-MUC at Leibniz Supercomputing Center (LRZ, www.lrz.de)(project number pn67ve). The authors gratefully acknowledge the computing time provided to them at the NHR Center NHR4CES at RWTH Aachen University (project number p0023446).
\footnotesize
\baselineskip 9pt

\clearpage
\thispagestyle{empty}
\bibliographystyle{proci}
\bibliography{FRASCINO_26}


\newpage

\small
\baselineskip 10pt


\end{document}


\acknowledgement{Supplementary material}\addvspace{10pt}

\acknowledgement{DNS initializaiton}\addvspace{10pt}

\noindent\textit{Velocity field definition}

\addvspace{10pt}

\begin{figure}[h!]
\renewcommand{\thefigure}{S\arabic{figure}}
\centering
\includegraphics[width=192pt]{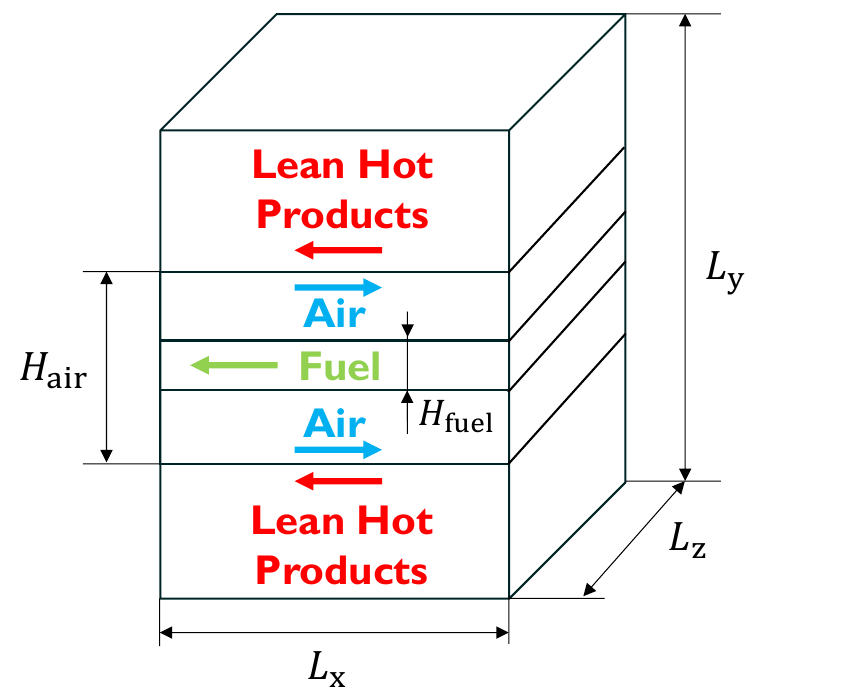}
\caption{\footnotesize Not-to-scale sketch of the numerical configuration.}
\label{sketch}
\end{figure}

The key parameters characterising the mixing layers for all four cases are summarised in Table~\ref{tab:dns_single} and the employed three-stream configuration (products--air--fuel--air--products) is depicted in Fig.~\ref{sketch}.
\begin{table}[h!]
\renewcommand{\thetable}{S\arabic{table}}
\setcounter{table}{0}
\centering
\caption{Key parameters for the four DNS cases.}
\label{tab:dns_single}
\resizebox{\columnwidth}{!}{
\begin{tabular}{lcccc}
\toprule
Case Name & \textbf{HD-FF} & \textbf{HD-SF} & \textbf{LD-FF} & \textbf{LD-SF} \\
\midrule
$H_{\mathrm{fuel}}$ [mm] & 0.8 & 0.8 & 3.9 & 3.9 \\
$H_{\mathrm{air}}$ [mm]  & 25 & 25 & 124 & 124 \\
$\Delta U_{\mathrm{FA}}$ [m/s] & 1.30 & 0.05 & 6.52 & 0.25 \\
$\Delta U_{\mathrm{HA}}$ [m/s] & 40   & 40   & 8.03 & 8.03 \\
$Re_{\mathrm{FA}}$             & 100  & 2    & 1129 & 43   \\
$Re_{\mathrm{HA}}$             & 10000 & 10000 & 10000 & 10000 \\
$Da_{\mathrm{FA}}$             & 0.2  & 5    & 0.2  & 5    \\
$Da_{\mathrm{HA}}$             & 0.2  & 0.2  & 5    & 5    \\
\bottomrule
\end{tabular}
}
\end{table}

The products--air mixing region is initialised from an instantaneous velocity field of a fully developed turbulent channel flow at $Re_{\tau} = 250$, computed prior to the DNS, following the approach of Attili et al.~\cite{Attili2014} and Scialabba et al.~\cite{Scialabba2025}. 

\begin{figure}[h!]
\renewcommand{\thefigure}{S\arabic{figure}}
\centering
\includegraphics[width=192pt]{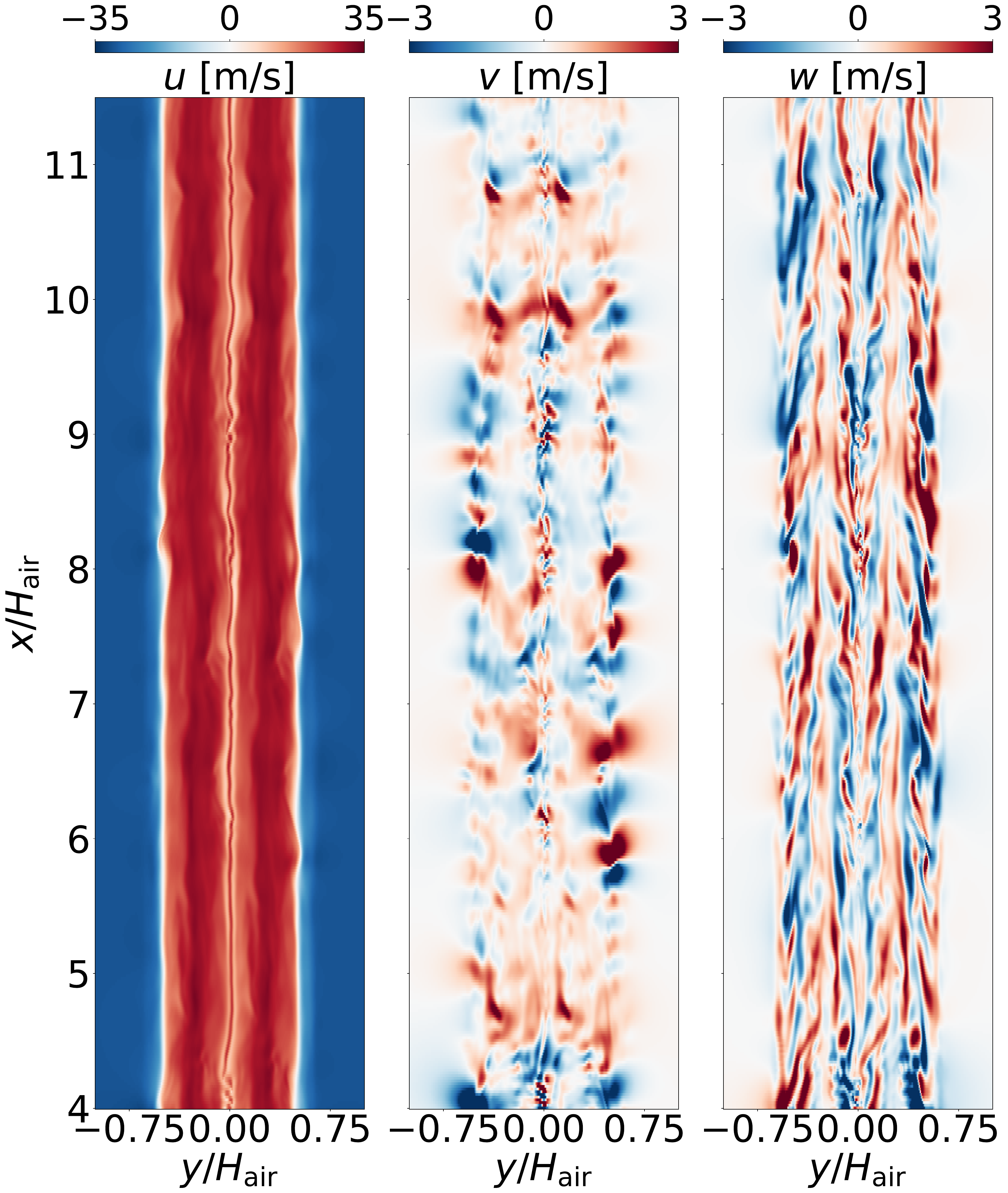}
\caption{\footnotesize Two-dimensional cuts of velocity components of the HD-FF case after initialization ($t = 0.5$ ms). Only a portion of the full domain ($L_\mathrm{x} = 30 H_\mathrm{{air}}$) in the streamwise direction x is shown.}
\label{fig:UVW_contours}
\end{figure}

The channel velocity field is uniformly rescaled to impose the prescribed velocity difference $\Delta U_{\mathrm{HA}}$ between the streams and interpolated in the region $(H_{\mathrm{fuel}}/2 \leq |y| \leq H_{\mathrm{air}}/2)$. Here, $\Delta U$ is defined as the difference between the bulk mean velocities of the two streams. The channel field is split into two halves and imposed on the hot products sides of the configuration ($(H_{\mathrm{air}}/2 \leq |y| \leq 3/4 \, H_{\mathrm{air}})$), while in the outer region ($y \geq |3/4 \, H_{\mathrm{air}}|$) a uniform velocity of $-U_{\mathrm{air}}$ is imposed. The fuel jet velocity profile $(y \leq |H_{\mathrm{fuel}}/2|)$ is instead prescribed as a parabolic profile, due to the low Reynolds numbers characterizing the fuel--air mixing layer across all cases (see Table~\ref{tab:dns_single}). The parabolic profile is scaled to match the prescribed velocity difference $\Delta U_{\mathrm{FA}}$ between the fuel and air streams. The velocity fields of the HD-FF case after initialization are shown in Fig.~\ref{fig:UVW_contours}. 

\begin{figure}[h!]
\renewcommand{\thefigure}{S\arabic{figure}}
\centering
\includegraphics[width=192pt]{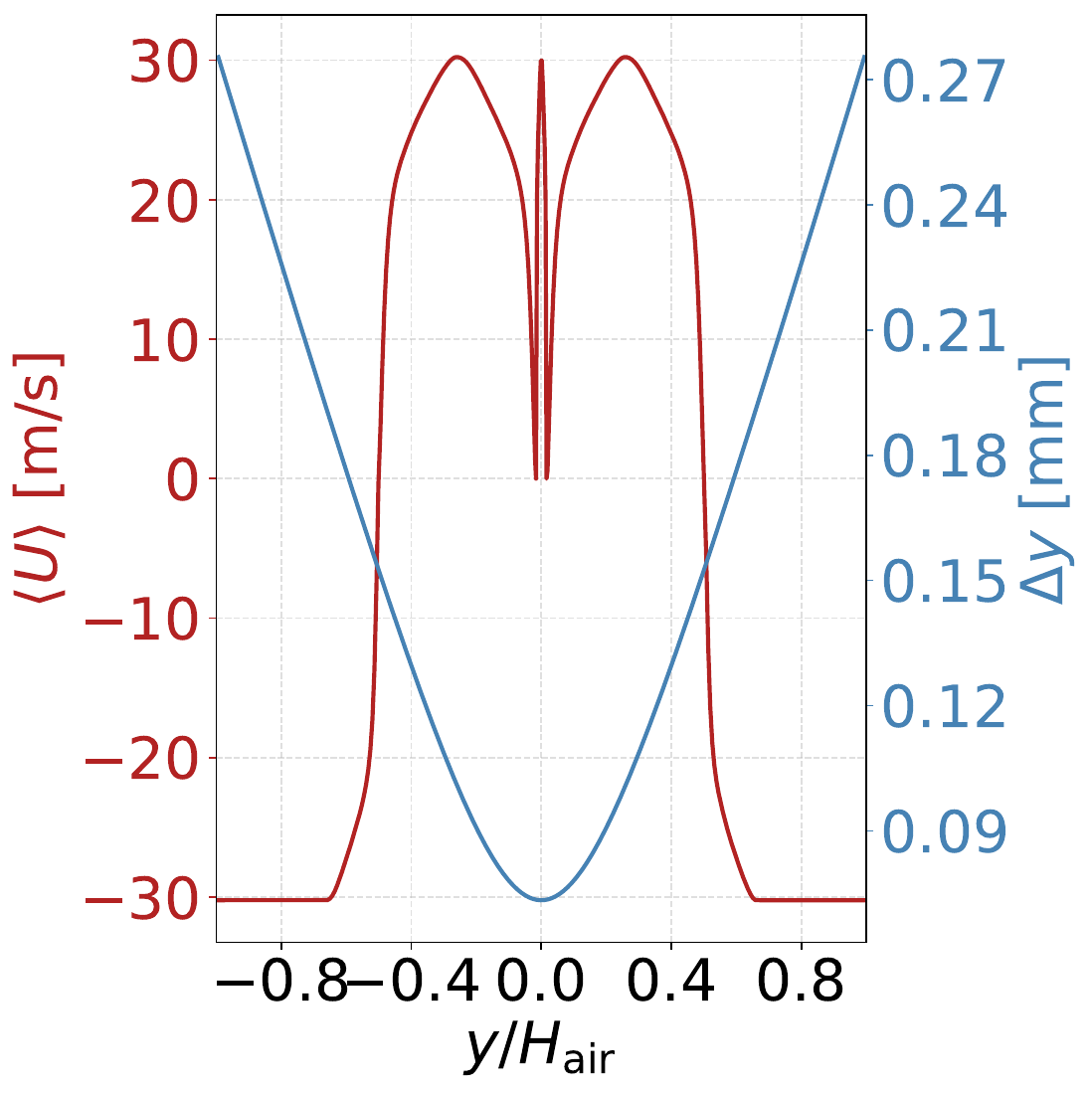}
\caption{\footnotesize Initial mean velocity profile (left) and spanwise grid spacing (right) for the HD-FF case.}
\label{fig:profiles}
\end{figure}

The computational grid in the spanwise direction y is generated using a hyperbolic-sine stretching function, which allows a smooth, monotone distribution of grid points with increased resolution in the region of interest near the fuel jet centerline.
Starting from a uniformly distributed parameter $\xi \in [-1,\,1]$, the descrete values are defined as
\begin{equation}
    \xi_\mathrm{i} = -1 + \frac{2(i-1)}{N_{\mathrm{y}}-1}\,, \qquad i = 1, \ldots, N_{\mathrm{y}}\,,
    \label{eq:csi}
\end{equation}
where $N_{\mathrm{y}}$ is the total number of points across the spanwise direction. The physical coordinate y is obtained through the transformation
\begin{equation}
    \hat{y}_\mathrm{i} = \frac{\sinh\!\left[\beta\,(\xi_\mathrm{i} - \xi_\mathrm{c})\right]}{\sinh(\beta)}\,,
    \label{eq:sinh_stretch}
\end{equation}
where $\beta$ is the stretching factor and $\xi_\mathrm{c}$ is the location of maximum refinement in the mapped space. The stretched coordinate $\hat{y}_\mathrm{i} \in [-1,\,1]$ is then linearly rescaled to the physical domain $[y_{\min},\,y_{\max}]$:
\begin{equation}
    y_\mathrm{i} = y_{\min} + \frac{(\hat{y}_\mathrm{i} + 1)}{2}\,(y_{\max} - y_{\min})\,.
    \label{eq:rescale}
\end{equation}
For the employed simulations, $\beta$ is equal to 4. This ensures a resolution of 22 grid points in the y-direction across the fuel jet. 
Fig.~\ref{fig:profiles} shows the mean velocity profiles and the corresponding variation of the grid spacing along the y-direction using Eqs.~\ref{eq:csi}, \ref{eq:sinh_stretch}, and \ref{eq:rescale}.

\addvspace{10pt}
\noindent\textit{Scalar field mapping}

\addvspace{10pt}

The fuel/air/products system is described as a ternary mixture using the following mixture fractions:
\begin{itemize}
\item $Z_{\mathrm{hot}}$: mixture fraction associated with the hot products stream. $Z_{\mathrm{hot}} = 1$ corresponds to pure hot products, while $Z_{\mathrm{hot}} = 0$ indicates the absence of hot products, i.e., a mixture of fuel and/or air.

\item $Z_{\mathrm{fuel}}$: mixture fraction associated with the fuel stream. $Z_{\mathrm{fuel}} = 1$ corresponds to pure fuel, while $Z_{\mathrm{fuel}} = 0$ indicates the absence of fuel, i.e., a mixture of air and/or hot products.

\item $Z_{\mathrm{air}} = 1 - Z_{\mathrm{fuel}} - Z_{\mathrm{hot}}$: complementary air mixture fraction. 
\end{itemize} 
The thermochemical states corresponding to the three pure streams are summarized in Table~\ref{tab:composition}.
\begin{table}[h!]
\centering
\caption{Compositions and thermodynamic conditions of the three streams.}
\label{tab:composition}
\resizebox{0.85\columnwidth}{!}{
\begin{tabular}{lccc}
\toprule
 & \textbf{$Z_\mathrm{hot} = 1$} & \textbf{$Z_\mathrm{fuel} = 1$} & \textbf{$Z_\mathrm{air} = 1$} \\
\midrule
$Y_{\mathrm{O_2}}$  & 0.0446 & ---    & 0.2330 \\
$Y_{\mathrm{CO_2}}$ & 0.0706 & ---    & ---    \\
$Y_{\mathrm{H_2O}}$ & 0.1445 & ---    & ---    \\
$Y_{\mathrm{N_2}}$  & 0.7403 & ---    & 0.7670 \\
$Y_{\mathrm{H_2}}$  & ---    & 0.0402 & ---    \\
$Y_{\mathrm{CH_4}}$ & ---    & 0.9598 & ---    \\
$T$ [K]             & 1225   & 300    & 900    \\
$p$ [atm]             & 1   & 1    & 1    \\
\bottomrule
\end{tabular}
}
\end{table}

\begin{figure}[h!]
\centering
\renewcommand{\thefigure}{S\arabic{figure}}
\includegraphics[width=192pt]{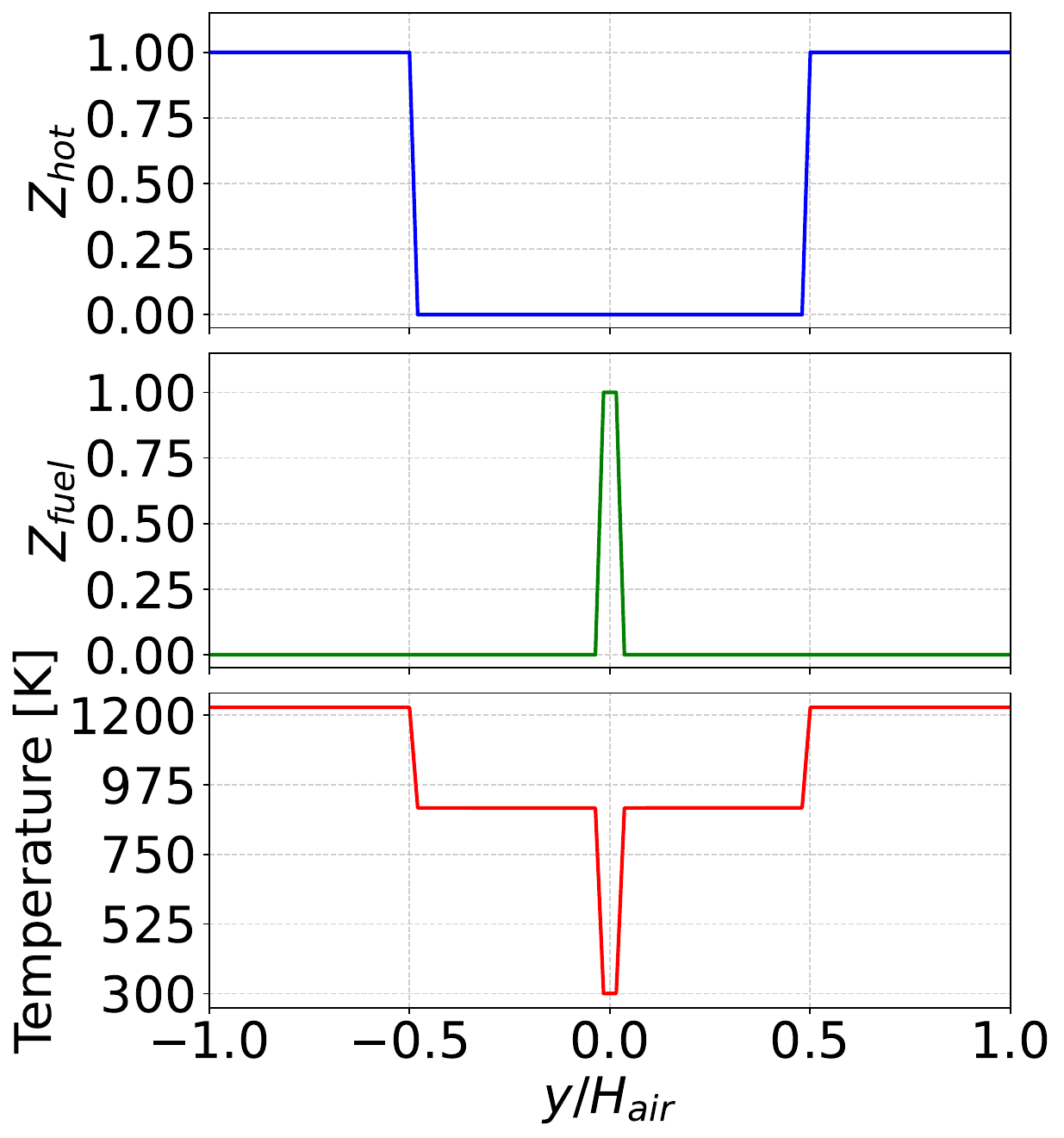}
\caption{\footnotesize Initial profiles of $Z_\mathrm{hot}$ (top), $Z_\mathrm{fuel}$ (middle), and the resulting mapped temperature (bottom) for the HD-FF case.}
\label{fig:ZZT}
\end{figure}

Both $Z_{\mathrm{hot}}$ and $Z_{\mathrm{fuel}}$ profiles are defined with the following linear function
\begin{equation}
Z(y) = \frac{|y| - (H/2 - w)}{2w}\,,
\end{equation}
where $H$ is the slab width ($H_{\mathrm{air}}$ for $Z_{\mathrm{hot}}$ and $H_{\mathrm{fuel}}$ for $Z_{\mathrm{fuel}}$), and $w$ is the transition layer thickness, equal to $0.5$ mm, with $Z=1$ and $Z=0$ outside the transition region. The obtained initial profiles are shown in Fig.~\ref{fig:ZZT} (top and middle).

The initial thermochemical state is obtained by mapping two extinguished flamelet solutions onto the established mixture fraction profiles. The first flamelet describes the mixing between hot products and air, while the second represents the mixing between fuel and air. The flamelets were computed with FlameMaster~\cite{FLAMEMASTER} using a reduced mechanism for lean methane-hydrogen blend combustion with 24 species and 251 reactions, which was derived from the full mechanism C3MechV3.3 model developed by Dong et al.~\cite {Dong2022}.
The thermochemical fields are initialized by applying each flamelet to its respective spatial region: outside the fuel-air mixing zone ($|y| \ge H_{\mathrm{fuel}}/2$), the state variables are mapped from the hot products-air flamelet according to the local value of $Z_{\mathrm{hot}}(y)$; in the central fuel slab ($|y| < H_{\mathrm{fuel}}/2$), the variables are mapped from the fuel-air flamelet using the local value of $Z_{\mathrm{fuel}}(y)$. 
The resulting temperature profile shown in Fig.~\ref{fig:profiles} (bottom) illustrates the combined effect of the two flamelet mappings and the consistency of the initialization procedure within the ternary mixture description.
\addvspace{10pt}

\begin{figure}[h!]
\renewcommand{\thefigure}{S\arabic{figure}}
\centering
\includegraphics[width=192pt]{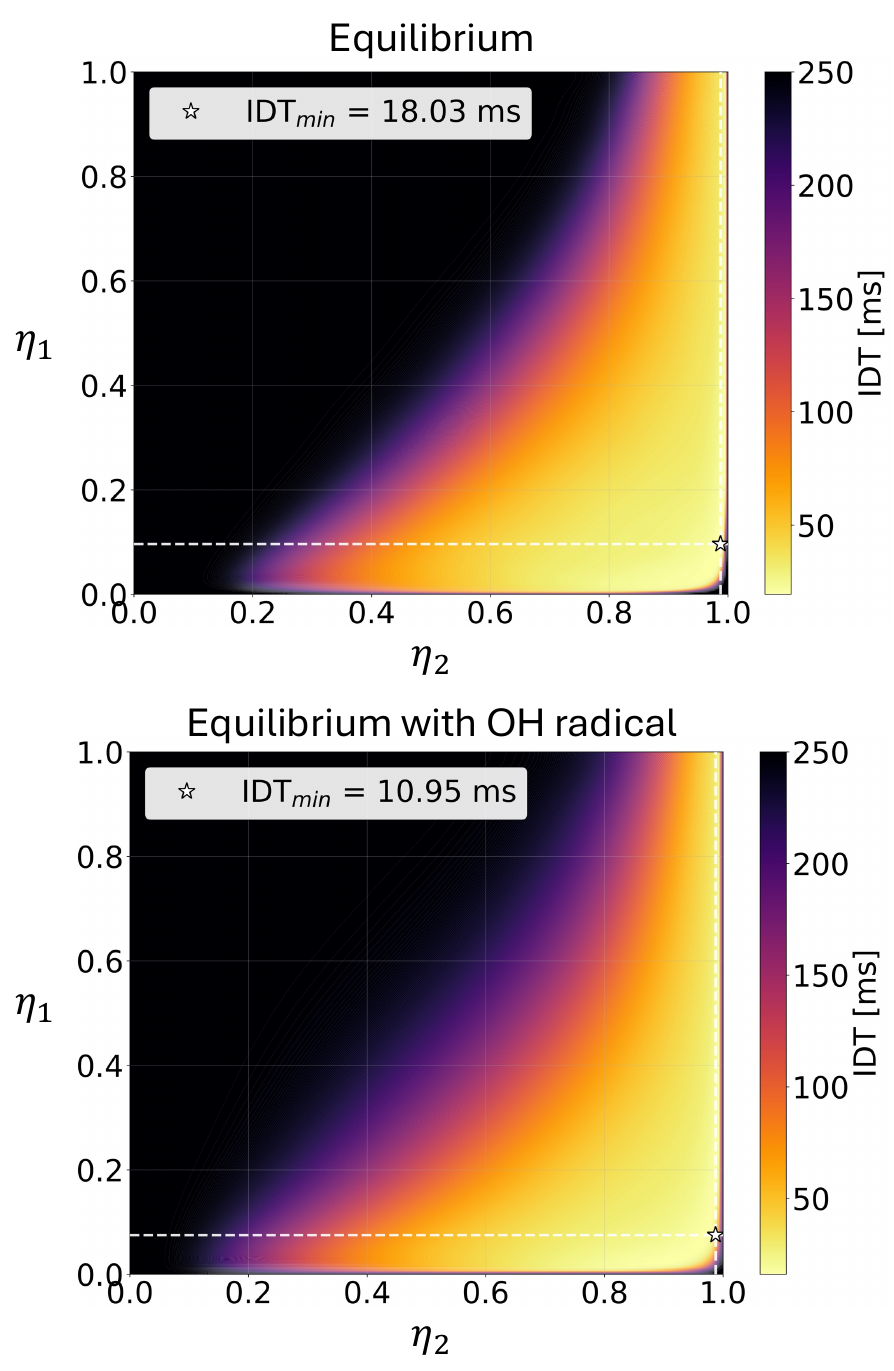}
\caption{Two-dimensional ignition delay time (IDT) maps as a function of the parametric coordinates $\eta_1 = Z_{\mathrm{hot}}$ and $\eta_2 = Z_{\mathrm{fuel}}/(1-Z_{\mathrm{hot}})$ for the hot products stream initialized with major species only (top) and with the addition of the OH radical (botom). The white dashed lines indicate the coordinates of the minimum ignition delay time.}
\label{fig:IDT_comparison}
\end{figure}

\acknowledgement{Influence of OH radical on the ignition delay time}
\addvspace{10pt}
To assess the influence of minor species in the hot products stream on the ignition delay time, an additional IDT map has been computed, including the OH radical in the hot products stream composition, alongside the major species considered in the main study. The comparison between the two cases is shown in Fig.~\ref{fig:IDT_comparison}. The inclusion of OH reduces the minimum ignition delay time from $\tau_{\mathrm{chem}} = \SI{18.03}{ms}$ to $\SI{10.95}{ms}$, a reduction of approximately 40\%, while the spatial structure of the IDT map and the location of the minimum in the mixture fraction space remain largely unaffected. This confirms that the radical pool primarily acts as an accelerator of the ignition process, without altering the qualitative structure of the ignition delay distribution. Consequently, the governing criterion based on $Da_{\mathrm{HA}}$ retains its validity: accounting for the radical pool would shift the MILD boundary in parameter space, as the reduced $\tau_{\mathrm{chem}}$ would require a proportional reduction of the mixing time, achievable, for instance, by reducing $H_{\mathrm{air}}$ or by increasing $\Delta U_{\mathrm{HA}}$, to maintain the same $Da_{\mathrm{HA}}$ value.

\begin{figure}[h!]
\renewcommand{\thefigure}{S\arabic{figure}}
\centering
\includegraphics[width=192pt]{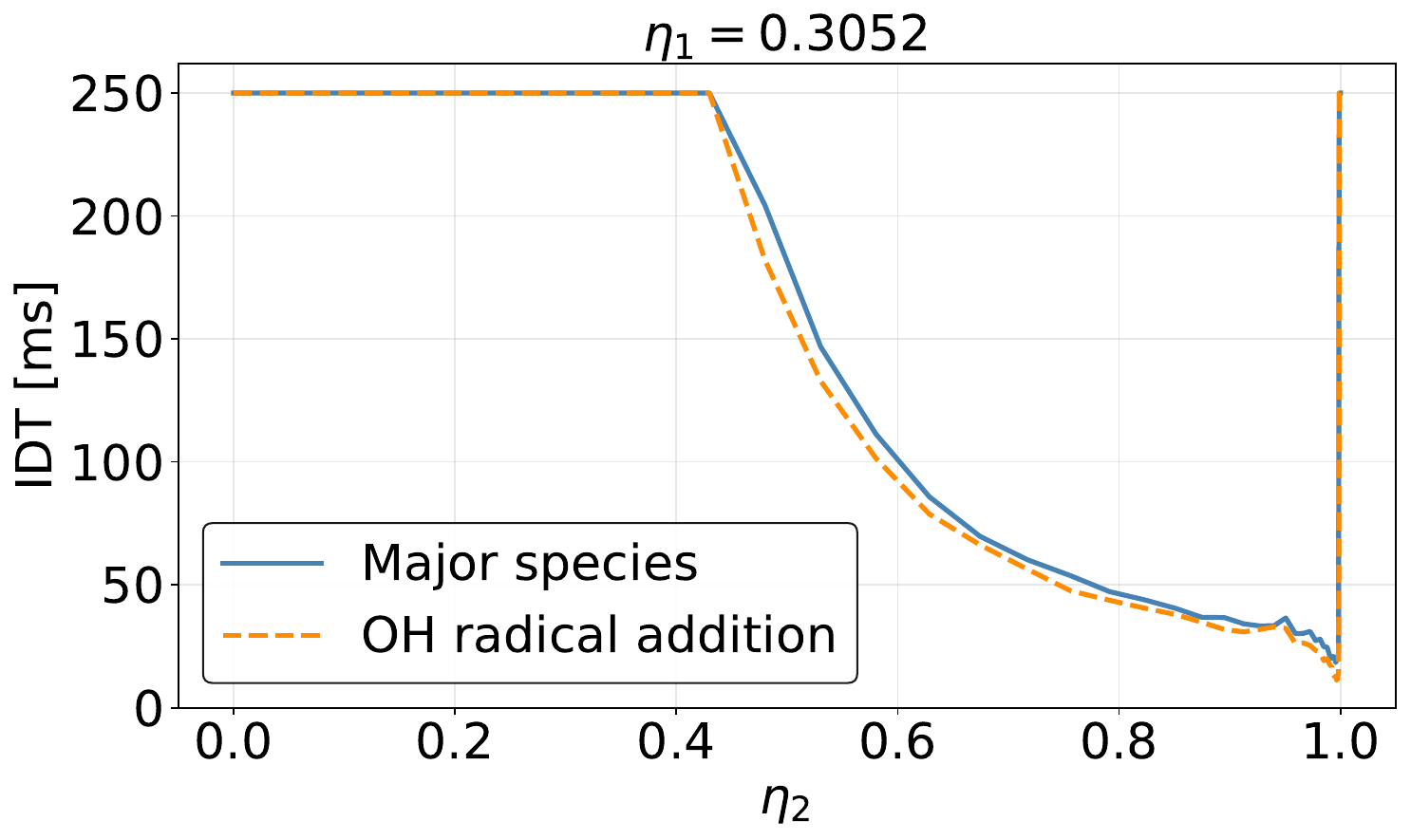}
\caption{One-dimensional IDT profiles as a function of $\eta_2$ at fixed $\eta_1 = 0.3052$, comparing the case with major species only and the case with the addition of the OH radical.}
\label{fig:IDT_slice}
\end{figure}

Additionally, a one-dimensional comparison of the IDT profiles as a function of $\eta_2$ at a fixed value of $\eta_1$ is provided in Fig.~\ref{fig:IDT_slice}, further highlighting that the inclusion of the OH radical uniformly reduces the ignition delay time across the entire range of $\eta_2$, while preserving the shape of the distribution and the location of the minimum.
\acknowledgement{CEMA: mathematical framework and computation}

\addvspace{10pt}
The chemical explosive mode analysis (CEMA) can be employed to characterize the local combustion dynamics. With such analysis, it can be defined whether the system evolves under autoignition conditions (0D-reactor-like behavior) or, instead, it exhibits deflagration-like behavior. The method has been implemented in the in-house finite-differences solver CIAO~\cite{Desjardins2008}.
Originally introduced by Lu et al.~\cite{CEMA_1} and later extended to include diffusion effects by Xu et al.~\cite{CEMA_2}, CEMA is based on the eigen-analysis of the Jacobian matrix $\mathbf{J}_\mathrm{\omega}$ of the chemical source term with respect to the thermochemical state vector $\boldsymbol{\psi} = (Y_\mathrm{1}, Y_\mathrm{2}, \ldots, Y_\mathrm{{N_\mathrm{s}}}, T)^\top$, where $Y_\mathrm{k}$ denotes the mass fraction of species $\mathrm{k}$ and $T$ the mixture temperature. The state vector has the dimension of $N_\mathrm{eq} = N_\mathrm{s} + 1$, where $N_\mathrm{s}$ is the number of transported (non-steady-state) species. Each element of the Jacobian is defined as
\begin{equation}
    J_\mathrm{ij} = \frac{\partial \dot{\omega}_\mathrm{i}}{\partial \psi_\mathrm{j}}\,,
\end{equation}
where $\dot{\omega}_\mathrm{i}$ is the source term of the $i$-th component of $\boldsymbol{\psi}$. The partial derivatives of the species source terms with respect to species mass fractions are precomputed using FlameMaster~\cite{FLAMEMASTER}, which provides the semi-analytic Jacobian entries directly from the chemical mechanism. The species source terms are expressed in terms of molar concentrations $[X_\mathrm{k}] = \rho Y_\mathrm{k} / W_\mathrm{k}$, so the Jacobian entries involving species are rescaled by the local density $\rho$ where appropriate. 
The sensitivity of each species production rate to temperature, $\partial \dot{\omega}_\mathrm{k} / \partial T$, is provided by the mechanism library and divided by $\rho$ to maintain consistency with the mass-fraction formulation. The source term for temperature (at constant pressure) is
\begin{equation}
    \dot{\omega}_\mathrm{T} = -\frac{1}{c_\mathrm{p}} \sum_{k=1}^{N_\mathrm{s}} h_\mathrm{k} \dot{\omega}_\mathrm{k}\,,
\end{equation}
where $h_\mathrm{k}$ is the specific enthalpy of 
species $k$ and $c_\mathrm{p}$ is the mixture heat capacity. 

Once the full Jacobian $\mathbf{J}_\mathrm{\omega} \in \mathbb{R}^{N_\mathrm{eq} \times N_\mathrm{eq}}$ is assembled, its eigenspectrum is computed via a standard real non-symmetric eigensolver (LAPACK \texttt{DGEEV}), which returns the complete set of right eigenvectors $\mathbf{v}_\mathrm{r}^{(m)}$, left eigenvectors $\mathbf{v}_\mathrm{l}^{(m)}$, and complex eigenvalues $\lambda^{(m)} = \lambda_\mathrm{r}^{(m)} + i\,\lambda_\mathrm{i}^{(m)}$, for $m = 1, \ldots, N_\mathrm{eq}$.

The chemical explosive mode (CEM) is identified as the eigenpair associated with the largest positive real part of the eigenvalue spectrum, i.e.,
\begin{equation}
    \lambda_\mathrm{exp} = \max_\mathrm{m} \left\{ \lambda_\mathrm{r}^{(m)} \right\}\,.
\end{equation}
%
%
Once the explosive eigenvalue $\lambda_\mathrm{exp}$ is identified, the corresponding left eigenvector $\mathbf{b} = \mathbf{v}_\mathrm{l}^{(e)}$ is extracted. The eigenvector is first $L_2$-normalized,
\begin{equation}
    \hat{\mathbf{b}} = \frac{\mathbf{b}}{\|\mathbf{b}\|_2}\,,
\end{equation}
and subjected to a sign convention: the component of $\hat{\mathbf{b}}$ with the largest absolute value (the dominant component, located at index $k^*$) is used to fix the orientation,
\begin{equation}
    \tilde{\mathbf{b}} = \frac{\hat{b}_\mathrm{{k^*}}}{|\hat{b}_\mathrm{{k^*}}|}\, \hat{\mathbf{b}}\,,
\end{equation}
so that the dominant component is always positive. This convention removes the arbitrary sign ambiguity inherent in eigenvector computation and ensures consistent physical interpretation across different spatial locations and time steps. The chemical contribution to the explosive mode is defined as the projection of the chemical source term vector $\dot{\boldsymbol{\omega}}$ onto the explosive mode direction $\tilde{\mathbf{b}}$:
\begin{equation}
\phi_\mathrm{\omega} = \tilde{\mathbf{b}} \cdot \dot{\boldsymbol{\omega}} = \left|\sum_{k=1}^{N_\mathrm{s}} \tilde{b}_\mathrm{k}\, \dot{\omega}_\mathrm{k} + \tilde{b}_\mathrm{T}\, \dot{\omega}_\mathrm{T}\right|,
\end{equation}
where $\dot{\omega}_\mathrm{k} = \dot{c}_\mathrm{k} W_\mathrm{k} / \rho$ is the mass-fraction source term of species $k$ (with $\dot{c}_\mathrm{k}$ the molar production rate). Physically, $\phi_\mathrm{\omega} > 0$ indicates that the chemical source term is actively driving the system along the explosive direction, consistent with heat release and radical production during ignition. The non-chemical contribution $\phi_\mathrm{s}$ quantifies the projection of the diffusive transport terms onto the same explosive mode:
\begin{equation}
    \phi_\mathrm{s} = \sum_{k=1}^{N_\mathrm{s}} \tilde{b}_\mathrm{k}\, \mathcal{D}_\mathrm{k} 
    + \tilde{b}_\mathrm{T}\, \mathcal{D}_\mathrm{T}\,,
\end{equation}
where $\mathcal{D}_\mathrm{k}$ is the diffusive flux divergence of species $k$ and $\mathcal{D}_\mathrm{T}$ is the corresponding thermal diffusion contribution to the temperature equation, defined respectively as:
\begin{equation}
    \mathcal{D}_\mathrm{k} = \frac{\nabla \cdot (\rho D_\mathrm{k} \nabla Y_\mathrm{k})}{W_\mathrm{k}}\,, 
    \qquad 
    \mathcal{D}_\mathrm{T} = \frac{\nabla \cdot (\lambda \nabla T)}{\rho \bar{c}_\mathrm{v}}\,.
\end{equation}
Species diffusion fluxes are computed from Fickian diffusion using species-specific diffusivity coefficients $D_\mathrm{k}$, with spatial gradients evaluated on the local computational mesh using interpolation and differentiation operators consistent with the flow solver. Once both contributions are available, the CEMA mode indicator is defined as
\begin{equation}
    \alpha = \frac{\phi_\mathrm{s}}{\phi_\mathrm{\omega}}\,,
\end{equation}
and it is evaluated only at locations where the explosive eigenvalue is positive ($\lambda_\mathrm{exp} > 0$), i.e., within the explosive region of the flow field. For visualization purposes, both contributions are normalized by the largest magnitude between $|\phi_\mathrm{s}|$ and $|\phi_\omega|$ at each point to facilitate comparison. The sign and magnitude of $\alpha$ provide a local characterization of the combustion regime:
\begin{itemize}
    \item $\alpha > 1$: assisted-ignition mode, dominated by diffusion (deflagration-like behavior);  
    \item $|\alpha| < 1$: autoignition mode, governed primarily by chemistry;  
    \item $\alpha < -1$: extinguishing mode, where diffusion suppresses chemistry.  
\end{itemize}
Outside the explosive region ($\lambda_\mathrm{exp} \leq 0$), both $\phi_\mathrm{\omega}$, $\phi_\mathrm{s}$, and $\alpha$ are set to zero, as the explosive mode analysis is not physically meaningful in the absence of a positive explosive eigenvalue.

\begin{figure}[h!]
\renewcommand{\thefigure}{S\arabic{figure}}
\centering
\includegraphics[width=192pt]{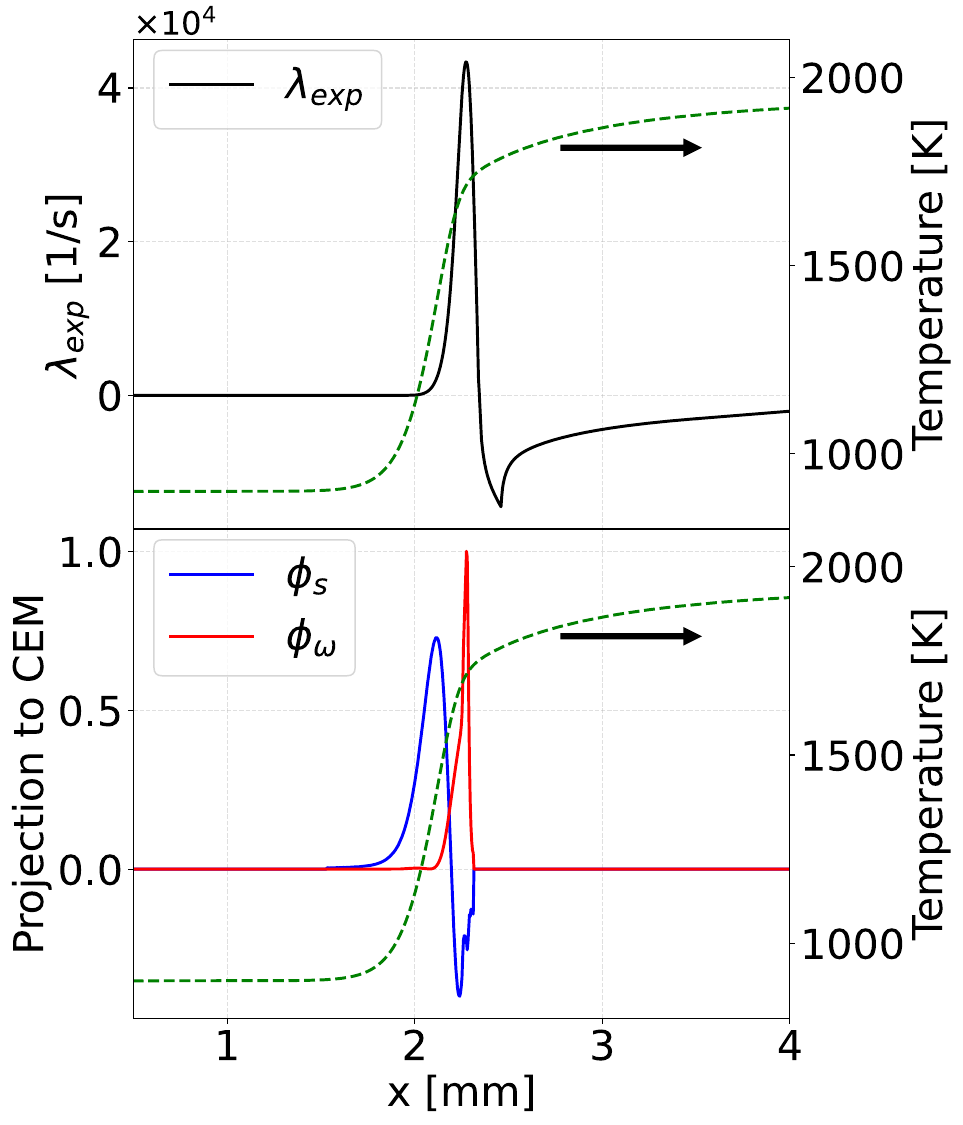}
\caption{\footnotesize 1D profiles of $\lambda_\mathrm{exp}$ (top) temperature, $\phi_{\omega}$ and $\phi_{\mathrm{s}}$ (bottom) for methane/air laminar premixed flame at an equivalence ratio of $\phi = 0.5$, under thermodynamic conditions of $T = 900\,\mathrm{K}$ and at atmospheric pressure.}
\label{fig:CEMA_VALIDATION}
\end{figure}

The validation of the developed model is performed using a 1D laminar, unstrained premixed methane-air flame at an equivalence ratio of $\phi = 0.5$, under thermodynamic conditions of $T = 900\,\mathrm{K}$ and atmospheric pressure. These conditions are the same employed in the validation of a similar framework from Gadalla~\cite{gadalla2022}. Figure~\ref{fig:CEMA_VALIDATION} shows the 1D profiles of $\lambda_\mathrm{exp}$, $\phi_{\mathrm{s}}$ and $\phi_{\omega}$, together with the temperature profile.

\addvspace{10pt}
\acknowledgement{CEMA and CSP: a comparative assessment}
\addvspace{10pt}

CEMA and Computational Singular Perturbation (CSP)~\cite{Lam1989,Goussis2021} are two closely related diagnostics rooted in dynamical systems theory; their mathematical connection and relative merits have been the subject of dedicated discussion in the literature~\cite{Goussis2021}. Both methods share the same mathematical foundation (the eigen-decomposition of the chemical Jacobian $\mathbf{J}_{\boldsymbol{\omega}}$) but differ in how the eigenspectrum is exploited. CEMA isolates the single mode with the largest positive real part of the eigenvalue, $\lambda_\mathrm{exp}$, and uses it to classify the local combustion regime. CSP, in contrast, considers the full ordered hierarchy of $N_\mathrm{eq}$ modes and constructs the Tangential Stretching Rate (TSR)~\cite{Valorani2015,Ciottoli2017},
\begin{equation}
  \mathrm{TSR} = \sum_{n=1}^{N_\mathrm{eq}} \mathrm{Re}(\lambda_n)\, A_n,
  \label{eq:tsr}
\end{equation}
where $A_n = \|\mathbf{f}^n\| / \sum_k \|\mathbf{f}^k\|$ is the normalised amplitude of the $n$-th CSP mode. The TSR is positive in autoignition-dominated regions ($\mathrm{TSR}>0$) and negative in deflagrative ones ($\mathrm{TSR}<0$), and is defined everywhere in the flow, including post-ignition zones where $\lambda_\mathrm{exp} \leq 0$ and CEMA is by construction silent. A known limitation of CEMA, discussed in detail by Goussis et al.~\cite{Goussis2021}, is the dormant-mode scenario: a small positive eigenvalue $\lambda_\mathrm{exp}$ may be formally present but dynamically negligible if its modal amplitude $A_n$ is vanishingly small compared to the dominant dissipative modes. In such cases, the TSR, being amplitude-weighted, correctly yields a negative (dissipative) value, whereas CEMA would report a positive explosive eigenvalue.

\begin{figure}[ht]
  \renewcommand{\thefigure}{S\arabic{figure}}
  \centering
  \includegraphics[width=\linewidth]{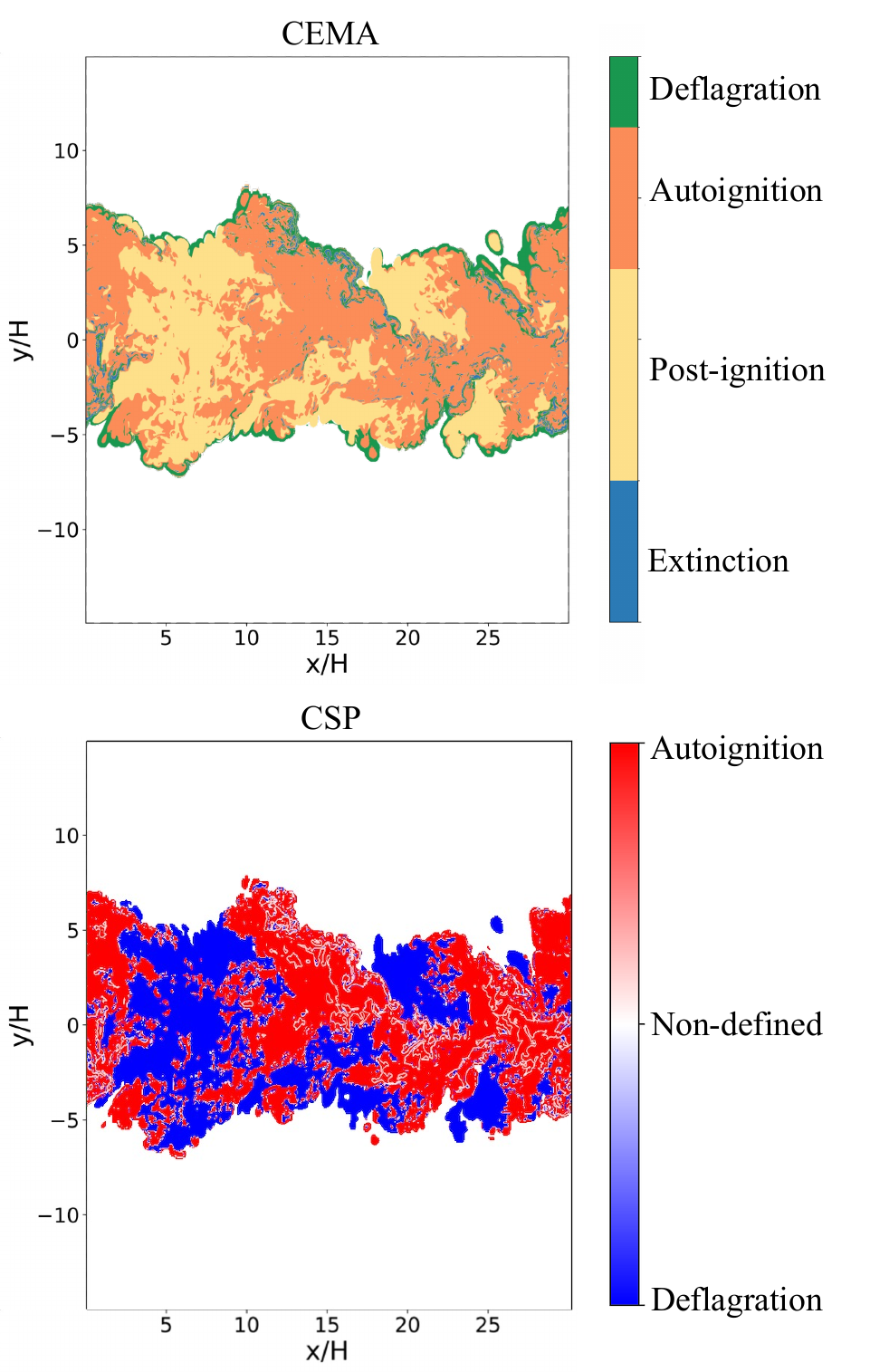}
  \caption{%
    Two-dimensional slices of the local combustion mode obtained from CEMA (top) and CSP/TSR (bottom) for the HD-FF case at $\tau_\mathrm{ign}$. CEMA colour map: orange = autoignition ($|\alpha|<1$), green = deflagration ($\alpha>1$), blue = extinction ($\alpha<-1$), yellow = post-ignition (CEMA undefined, $\lambda_\mathrm{exp} \leq 0$). CSP colour map: red = autoignition ($\mathrm{TSR}>0$), blue = deflagration ($\mathrm{TSR}<0$).
  }
  \label{fig:cema_csp}
\end{figure}

To assess whether this limitation affects the conclusions of the present study, both CEMA and CSP/TSR were applied to the DNS dataset and the resulting combustion mode maps were compared. Figure~\ref{fig:cema_csp} shows two-dimensional slices of the local mode indicator for the HD-FF case at the ignition time $\tau_\mathrm{ign}$.

For consistency with the CEMA formulation, the convective flux contribution was omitted from the TSR computation; the vector field was therefore taken as $\mathbf{g} = \dot{\boldsymbol{\omega}} + \mathbf{d}$, where $\mathbf{d}$ is the diffusive transport vector.

Two conclusions emerge from the comparison. First, the autoignition-dominated regions identified by CEMA ($|\alpha|<1$) and by CSP/TSR ($\mathrm{TSR}>0$) are spatially coincident, demonstrating that the dormant-mode scenario does not materialise in the pre-ignition zones that constitute the focus of the present analysis. Second, the two methods diverge in the post-ignition region: CEMA assigns these zones to the undefined category ($\lambda_\mathrm{exp} \leq 0$), while CSP/TSR classifies them as dissipative ($\mathrm{TSR}<0$). This reflects an intrinsic difference in scope: CEMA is an ignition-focused diagnostic, whereas the TSR is a global dynamical descriptor. The two methods are therefore complementary rather than contradictory.
The adoption of CEMA in the present study is then justified since the analysis is focused on the pre-ignition and early-ignition regions, where, as shown above, CEMA and CSP/TSR yield almost identical classifications.

\begin{figure}[h!]
\renewcommand{\thefigure}{S\arabic{figure}}
\centering
\includegraphics[width=192pt]{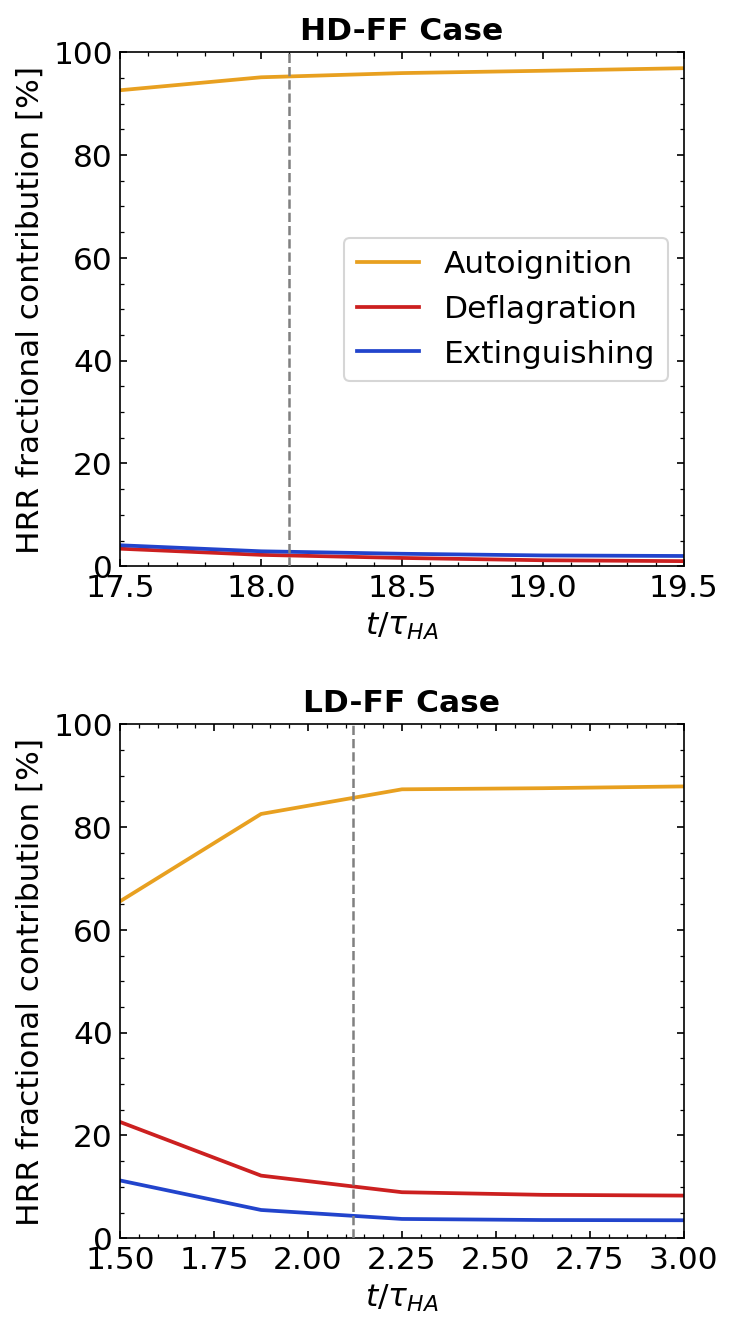}
\caption{Temporal evolution of the fractional HRR contributions of autoignition (orange), deflagration (red), and extinguishing (blue) modes for the HD-FF (top) and LD-FF (bottom) cases. The dashed vertical line marks the ignition time $\tau_{\mathrm{ign}}$ of each respective case.}
\label{fig:CEMA_time}
\end{figure}

\begin{figure}[h]
    \renewcommand{\thefigure}{S\arabic{figure}}
    \centering    
    \includegraphics[width=192pt]{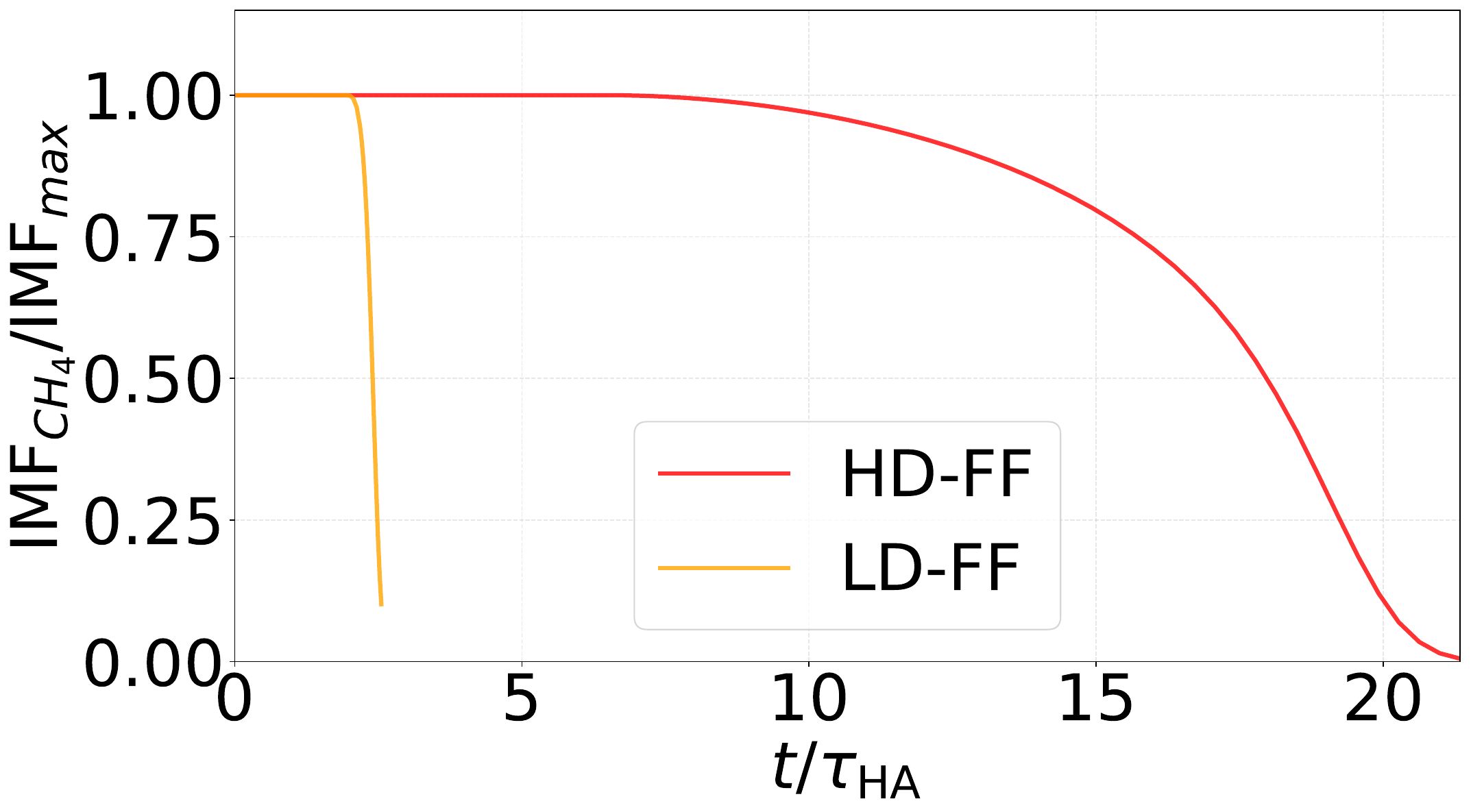}
    \caption{Temporal evolution of the normalized integral mass fraction of CH$_4$, $\mathrm{IMF}_{\mathrm{CH_4}}/\mathrm{IMF}_{\mathrm{max}}$, as a function of normalized time $t/\tau_{\mathrm{HA}}$ for the HD-FF and LD-FF cases.}
    \label{fig:IMF_CH4}
\end{figure}

\begin{figure}[h]
    \renewcommand{\thefigure}{S\arabic{figure}}
    \centering
    \includegraphics[width=\linewidth]{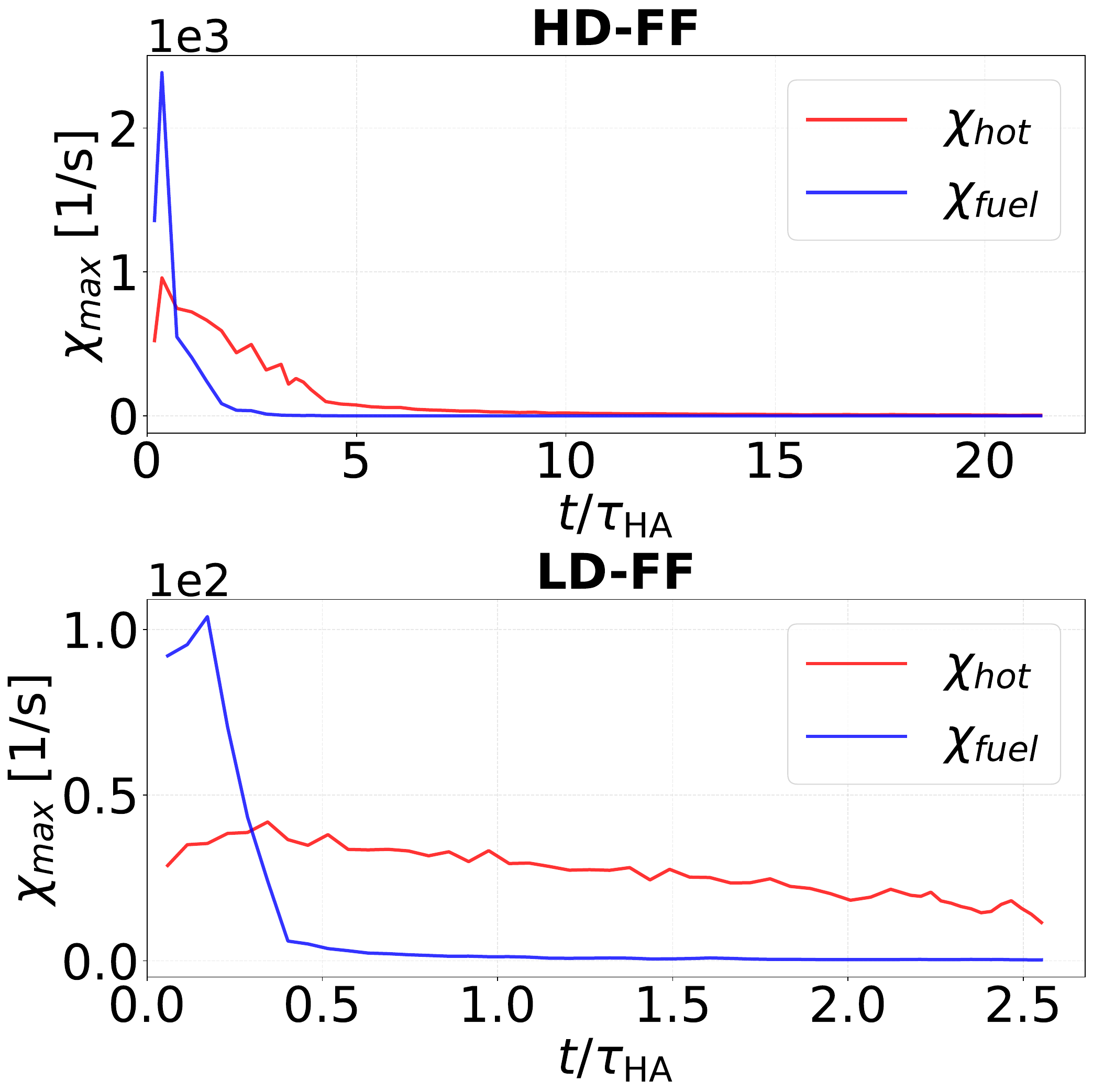}
    \caption{Maximum scalar dissipation rate $\chi_{\mathrm{max}}$ as a function of normalised time $t/\tau_{\mathrm{HA}}$ for the HD-FF (top) and LD-FF (bottom) cases. Red and blue lines denote $\chi_{hot}$ and $\chi_{fuel}$, respectively.}
    \label{fig:chi_max_vs_time}
\end{figure}

\addvspace{10pt}
\acknowledgement{Temporal evolution of the combustion mode HRR contributions, fuel consumption, and scalar dissipation rate}
\addvspace{10pt}
To assess whether the fractional heat release rate (HRR) contributions reported in Table~2 of the main manuscript are representative of the entire combustion process, the contributions of the autoignition, deflagration, and extinguishing modes have been tracked throughout the post-ignition oxidation phase for the representative HD-FF and LD-FF cases. The results are shown in Fig.~\ref{fig:CEMA_time}, where the dashed vertical line marks the ignition time $\tau_{\mathrm{ign}}$ of each respective case. The fractional HRR contributions of each combustion mode remain qualitatively consistent throughout the post-ignition oxidation phase, confirming that the values reported in Table~2 are representative of the combustion mode distribution over the entire oxidation process. The temporal evolution of the normalized integral mass fraction of CH$_4$, $\mathrm{IMF}_{\mathrm{CH_4}}/\mathrm{IMF}_{\mathrm{max}}$, is shown in Fig.~\ref{fig:IMF_CH4} as a function of the normalized time $t/\tau_{\mathrm{HA}}$ for the HD-FF and LD-FF cases. The HD-FF case shows a monotonic decrease to zero, confirming complete fuel consumption. For the LD-FF case, a minor residual CH$_4$ fraction is observed at the end of the simulation, attributed to locally fuel-rich regions arising from the non-homogeneous mixing conditions characteristic of this configuration, as discussed in Section~3.2. Finally, the temporal evolution of both scalar dissipation rates is reported in Fig.~\ref{fig:chi_max_vs_time}.

\addvspace{10pt}

\acknowledgement{Ignition progress conditioned on the cumulative scalar dissipation rate}

\begin{figure*}[h]
    \renewcommand{\thefigure}{S\arabic{figure}}
    \centering
    \includegraphics[width=\textwidth]{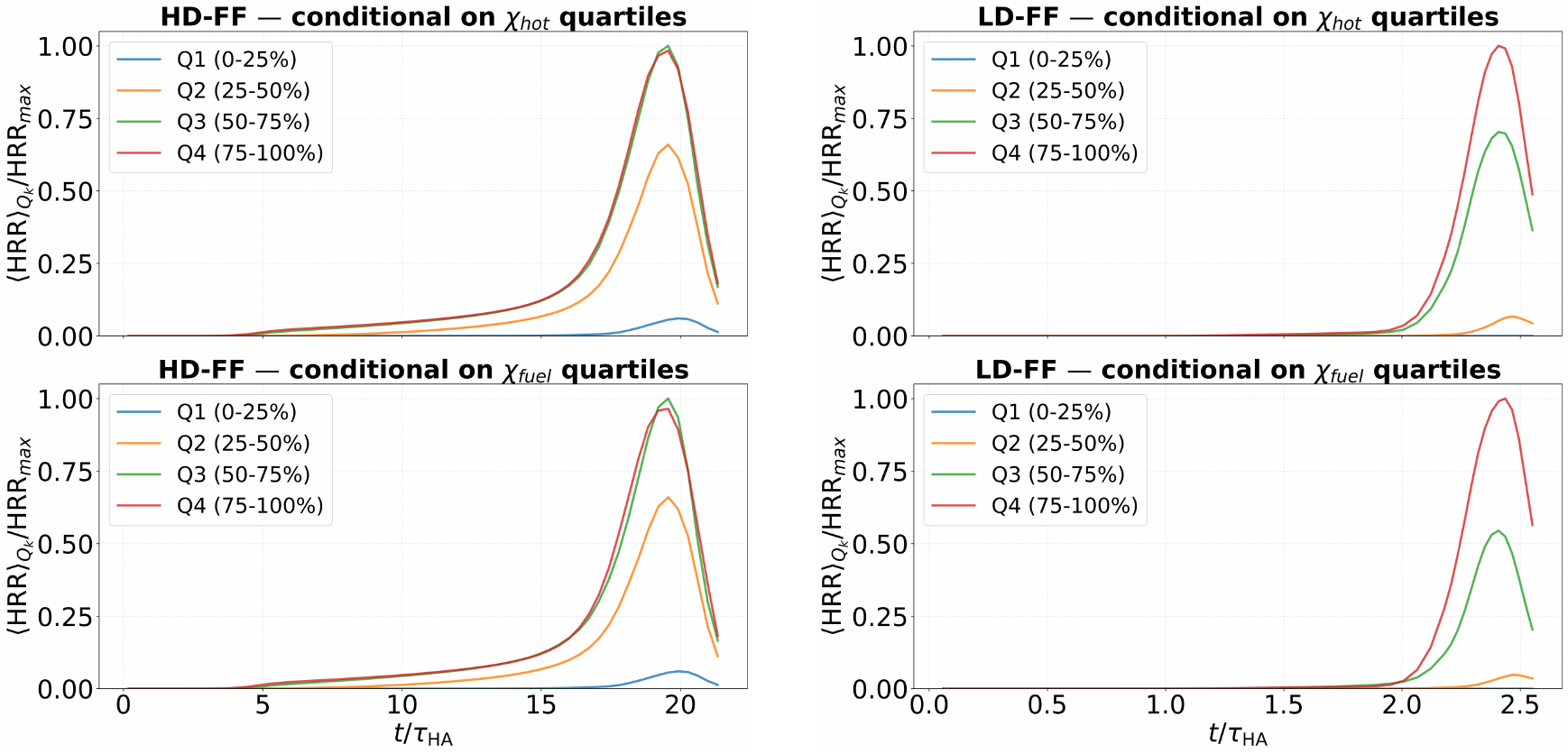}
    \caption{Conditionally averaged heat release rate $\langle \mathrm{HRR} \rangle_{Q_k} / \mathrm{HRR}_{\max}$ as a function of normalised time $t/\tau_{\mathrm{HA}}$, conditioned on fixed quartiles $Q_k$ ($k=1,\ldots,4$) of the cumulative scalar dissipation rate $\chi_{cum}(t)=\int_0^t\chi\,\mathrm{d}t'$. Left column: HD-FF; right column: LD-FF. Top row: quartiles based on $\chi_{hot}$; bottom row: quartiles based on $\chi_{fuel}$. $Q_1$--$Q_4$ denote the 0--25\%, 25--50\%, 50--75\% and 75--100\% percentile bins of $\chi_{cum}$, respectively.}
    \label{fig:S_HRR_quartiles}
\end{figure*}

To quantify the sensitivity of ignition to the cumulative scalar dissipation rate, we define
\begin{equation}
    \chi_{cum}(\mathbf{x},\, t) = \int_0^t \chi(\mathbf{x},\, t')\,\mathrm{d}t',
    \label{eq:chi_cum}
\end{equation}
which represents the total scalar dissipation experienced by each fluid element up to time $t$. The computational domain is partitioned into four equal-population quartiles $Q_1$--$Q_4$ based on the value of $\chi_{cum}$ at the final simulated time step. This procedure is performed independently for $\chi_{hot}$ and $\chi_{fuel}$, yielding two separate sets of quartile maps. Fig.~\ref{fig:S_HRR_quartiles} shows the conditionally averaged heat release rate $\langle\mathrm{HRR}\rangle_{Q_k}$, normalised by the instantaneous global maximum $\mathrm{HRR}_{\max}$, for both cases and both scalar dissipation rate definitions. In both HD-FF and LD-FF, a clear and monotonic hierarchy is observed: fluid elements in the upper quartiles ($Q_3$, $Q_4$), which have experienced the highest cumulative scalar dissipation, ignite earlier and sustain significantly higher heat release rates than those in the lower quartiles. The lowest quartile $Q_1$ remains virtually unreacted throughout the simulation in both cases, demonstrating that regions of persistently low mixing activity do not contribute meaningfully to the combustion process. The separation between quartiles is particularly pronounced in the LD-FF case, where the weaker jet momentum produces a broader spatial distribution of $\chi_{cum}$, amplifying the differences in ignition behaviour across the domain. These results confirm that ignition is strongly selective and controlled by the local history of scalar dissipation, consistent with the classical sensitivity of autoignition to mixing rate fluctuations~\cite{Mastorakos2009}.
\thispagestyle{empty}
\bibliographystyle{proci}
\bibliography{FRASCINO_26_supplementary}
 
\clearpage
\begin{figure*}[h!]
\renewcommand{\thefigure}{S\arabic{figure}}
\centering
\includegraphics[width=\textwidth]{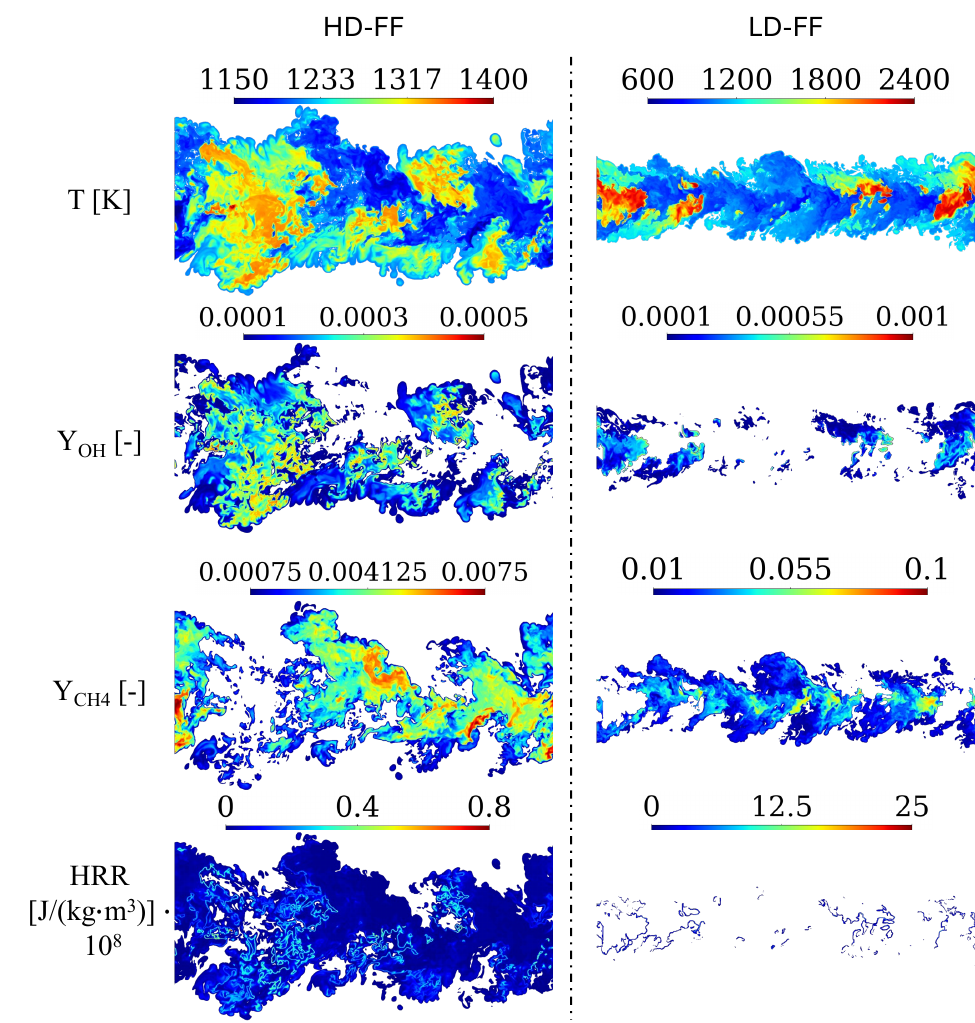}
\caption{Two-dimensional slices of temperature $T$, OH mass fraction $Y_{\mathrm{OH}}$, CH$_4$ mass fraction $Y_{\mathrm{CH_4}}$, and heat release rate HRR for the HD-FF (left) and LD-FF (right) cases at their respective ignition times $\tau_{\mathrm{ign}}$.}
\label{fig:cuts_end}
\end{figure*}
\addvspace{10pt}

\begin{figure*}[h!]
\renewcommand{\thefigure}{S\arabic{figure}}
\centering
\vspace{-0.4 in}
\includegraphics[width=\textwidth]{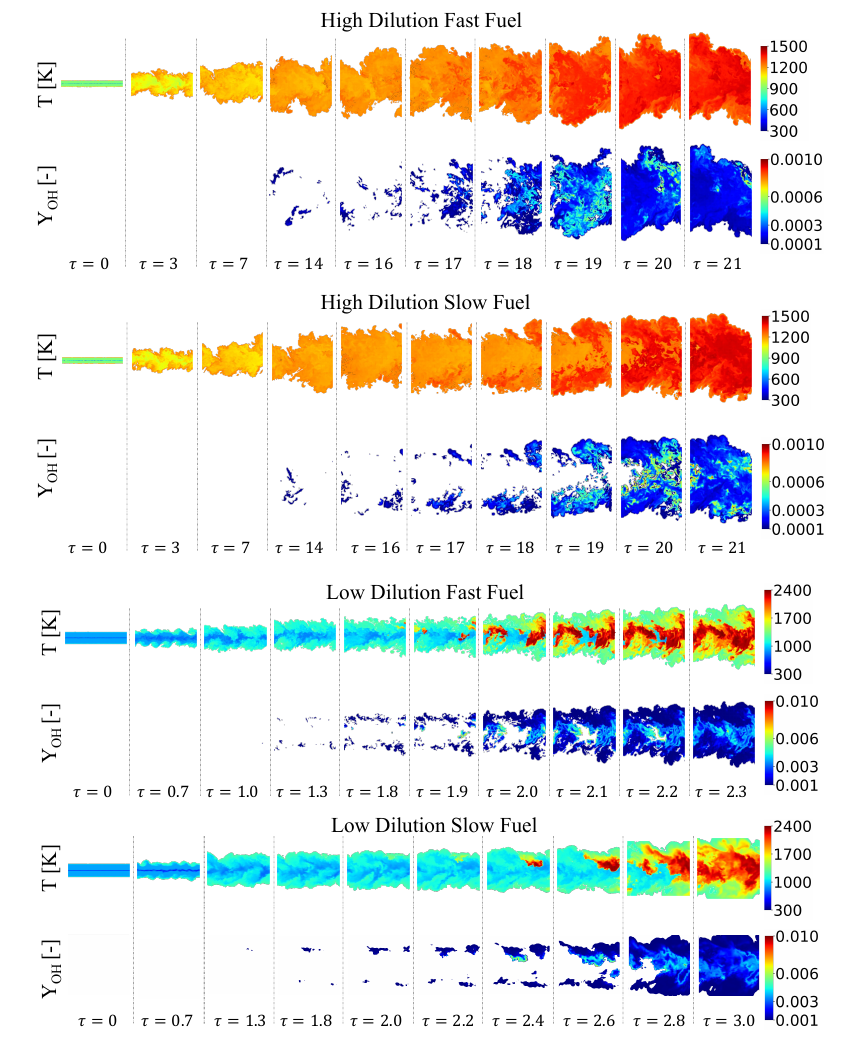}
\caption{Two-dimensional temperature and OH mass fraction field slices for all simulated cases at representative time instants. The reference time $\tau$ shown below each slice corresponds to the ratio between the physical time $t$ and the respective case mixing time ($\tau_{\mathrm{HA}}$).}
\label{HDFF_slices}
\end{figure*}

\footnotesize
\baselineskip 9pt



\newpage

\small
\baselineskip 10pt
